\documentclass[CRPHYS,Unicode,manuscript]{cedram}

\title{Reflections on dipolar quantum fluids}

\alttitle{R\'{e}flexions sur les fluides quantiques dipolaires}

\author{\firstname{Wilhelm} \lastname{Zwerger}}
\address{Physik Department TU M\"unchen, James Franck Strasse, 85748 Garching,
Germany}
\email[W. Zwerger]{zwerger@tum.de}

\keywords{Dipolar fluids }

\begin{abstract}
We present a thermodynamic description of ultracold gases with dipolar interactions
which properly accounts for the long-range nature and broken 
rotation invariance of the interactions. It involves an additional thermodynamic 
field conjugate to the linear extension of the gas along the direction of the dipoles.
The associated uniaxial pressure shows up as a deviation 
from the Gibbs-Duhem relation in the density profile of a trapped gas. It has to 
vanish in self-bound droplets, a condition which determines the observed dependence
of the aspect ratio on particle number. A tensorial generalization 
of the virial theorem and a number of further exact thermodynamic relations are derived.  
Finally, extending a model due to Nozi\`eres, a simple criterion for the freezing transition
to a superfluid mass density wave is given.  
 \end{abstract}

\begin{altabstract}
  Résumé en français.
\end{altabstract}

\begin{document}

\maketitle

\section{Introduction}

The realization of a Bose Einstein condensate with Chromium~\cite{grie05} has opened a new field of 
research in ultracold gases. It allows to explore a wide range of phenomena which are uniquely tied to
the long-range and partially attractive nature of dipolar interactions. Following the extension to condensates 
with Erbium or Dysprosium where the strength of these interactions may exceed the short-range repulsion, 
 the field has grown immensely in recent years. This is based, in particular, on the discovery 
of self-bound droplets~\cite{ferr16} and of supersolid phases~\cite{boet19,tanz19a,chom19}, where superfluidity
coexists with broken translation invariance: see Ref.~\cite{chom23} for a recent review of magnetic dipolar 
gases and the Lectures~\cite{zwer21,dali24} for an introduction to the underlying concepts. On the theory 
side, a successful qualitative description of the 
observations is provided by an extended Gross-Pitaevskii equation, where a non-analytic contribution 
proportional to $\vert\psi\vert^5$ is added to the energy functional. As will be discussed below, a
proper microscopic derivation of this procedure in the relevant regime of strong dipolar interations, where
the chemical potential may assume negative values, is lacking. It is therefore of interest to develop
a description which only relies on general thermodynamic relations, properly accounting
for the long-range and anisotropic nature of the interactions. A step in this direction will be taken in 
the present contribution for quantum fluids in thermal equilibrium,
extending concepts from classical polar fluids~\cite{hans06} and liquid crystals~\cite{chai95}. 
In particular, it is shown that the anisotropy and long-range nature
of the interactions require the introduction of an additional contribution $h_1 dL_z$ in the 
differential of the free energy, with $L_z$ the system length along the direction of the dipoles.
It gives rise to the shape-dependence of thermodynamic properties and 
an anisotropy in the momentum current tensor $\underline{\Pi}(x)$ and thus effectively to pressure.  
The appearance of an additional extensive variable $L_z$ in the free energy $F(T,V,L_z,N)$ beyond particle 
number and volume leads to a violation of the Gibbs-Duhem relation which is reflected 
in the density profile of harmonically trapped gases. Moreover, for the case of self-bound droplets
where the internal forces have to balance locally, the vanishing of the uniaxial contribution implies
a scale-dependent aspect ratio $\kappa(N)\sim N^{-1/4}$, approaching zero
in the thermodynamic limit where the droplets evolve into a needle-like shape. Finally, we 
address the nature of the transition from a homogeneous superfluid to a supersolid and discuss
a simple model due to Nozi\`eres~\cite{nozi04} which provides a criterion for the associated 
critical value of the roton gap beyond mean-field.\\

In general, the aim of the present contribution is to clarify a number of conceptual points in
the theory of dipolar gases, thus providing a better understanding of some fundamental issues 
and open problems.

\section{Thermodynamics of uniaxial quantum fluids}

In close analogy to the standard description of classical polar fluids~\cite{hans06}, 
the two-body interaction
\begin{equation}
V(x_{12})=V_{\rm sr}(r_{12})-\frac{\mu_m^2}{4\pi}\,\frac{2\,P_2(\hat{z}\cdot\hat{x}_{12})}{r_{12}^3}\cdot\chi(r_{12})
 \label{eq:V12}
 \end{equation} 
in Bose quantum fluids whose permanent dipoles $\mu_m$ are all oriented along the $z$-direction
may be separated into a rotation invariant short-range plus the long-range, anisotropic dipolar contribution
(we use magnetic dipoles in the following and units where $\mu_0\equiv 1$ but our results also hold for 
electric ones with minor changes).  
Here, $P_2(x)=(3x^2-1)/2$ is the standard Legendre polynomial and $\hat{z}\, ,\,\hat{x}_{12}$ are unit vectors 
along, respectively, the dipole orientation or the separation vector $x_{12}=x_1-x_2$. In the ultracold limit, the 
two-body scattering amplitude associated with $V_{\rm sr}(r_{12})$ only involves s-wave scattering and thus is 
fully characterized by the scattering length $a$, which is assumed to be positive. The short-range interaction 
can therefore be replaced by a zero-range pseudopotential with strength parameter $g=4\pi\hbar^2 a/m$.
A similar simplification is possible for the dipolar interaction, whose strength defines an additional
effective length $a_{\rm dd}>0$ by $g_{\rm dd}=\mu_m^2/3=4\pi\hbar^2 a_{\rm dd}/m$. Indeed, dipolar scattering at 
low energies arises from large separations, where $V_{\rm dd}(x_{12})\sim \mu_m^2/r_{12}^3$ is weak. 
Asymptotically, the two-body scattering amplitude
\begin{equation}
f_{\rm dd}(\hat{k}_i\to\hat{k}_f, |k|)\; \xrightarrow[|k|a_{\rm dd}\ll 1]\; f_{\rm dd}^{(B)}(\hat{k}_i\to\hat{k}_f)=-a_{\rm dd}\cdot 2 P_2(\hat{z}\cdot\hat{q})
\hspace*{0.2cm} {\rm for} \hspace*{0.2cm} |q|=2|k|\sin{(\theta/2)}\ne 0
\label{eq:Born}
 \end{equation} 
is therefore given by the Born approximation, proportional to the Fourier transform $V_{\rm dd}(q)=g_{\rm dd}\, 2 P_2(\hat{z}\cdot\hat{q})$
\nobreak of the bare dipolar interaction at the momentum transfer wave vector $q=k_f-k_i$. Note that $V_{\rm dd}(q)$ is
independent of $|q|$ only for $|q|r_0\ll 1$, where the short distance cutoff function $\chi(r)$ in Eq.~\eqref{eq:V12} 
may be replaced by unity and also that the forward scattering limit $\theta=0$ is ill-defined. Moreover, since
$V_{\rm dd}(x_{12})$ is not rotational symmetric, the scattering amplitude cannot be decomposed into separate 
angular momenta and also depends on the initial direction $\hat{k}_i$ 
\footnote{A discussion of how this affects rethermalization in dipolar gases has been given by Bohn and Jin~\cite{bohn14}.}.
The validity of the result ~\eqref{eq:Born} 
has been tested by Bohn et. al.~\cite{bohn09} based on a numerical solution of the full dipolar two-body
scattering problem. Using a pure power law dependence of $V_{\rm dd}(x_{12})$ with a sharp cutoff at a 
distance $r_0\ll a_{\rm dd}$, the cross section $\bar{\sigma}(|k|)$ averaged over all incident directions $\hat{k}_i$
is found to essentially coincide with the Born approximation value obtained for $|k|a_{\rm dd}\to 0$ up to energies
of order $\hbar^2/(ma_{\rm dd}^2)$. 
For typical values $a_{\rm dd}\approx a \simeq 100\, a_0$,
dipolar scattering in ultracold gases is therefore fully described in terms of a single parameter $a_{\rm dd}$ and 
deviations of the actual dipolar interaction from a pure power law at short distances $r_{12}< r_0$ are not relevant.
This standard assumption, where point-like dipoles are combinded with a zero-range pseudopotential
has to be used with caution, however. Indeed, the cutoff scale $r_0$ explicitely enters the corrections 
beyond the Born approximation. Specifically, while both the real and the imaginary part ${\rm Im} f_{\rm dd}(0)=|k|\,\sigma_{\rm dd}(\hat{k}_i)/(4\pi)\sim |k| a_{\rm dd}^2$ of 
the forward scattering amplitude vanish at low energies and are independent of $r_0$, the real part ${\rm Re} f_{\rm dd}^{(2\,B)}(0)\sim a_{\rm dd}^2/r_0$
is finite in second order Born approximation and linearly sensitive to the short distance cutoff.  
A more serious problem arises at the many-body level where the expectation value of the dipolar interaction 
\begin{equation}
\langle\hat{H}_{\rm dd}\rangle = \int_{x_1}\int_{x_2}\left[-\frac{\mu_m^2}{8\pi}\,\frac{2\,P_2(\hat{z}\cdot\hat{x}_{12})}{r_{12}^3}\right]\,\rho^{(2)}(x_1,x_2)
\hspace*{0.2cm} {\rm with} \hspace*{0.2cm} \rho^{(2)}(x_1,x_2)=\rho^{(1)}(x_1)\rho^{(1)}(x_2)\cdot g^{(2)}(x_1,x_2)
 \label{eq:Hdd}
 \end{equation} 
depends on the exact two-particle density $\rho^{(2)}(x_1,x_2)$. It thus requires knowledge not only of the
one-particle density $\rho^{(1)}(x)$ of the atomic cloud but also of the associated two-particle distribution 
function $g^{(2)}(x_1,x_2)$ (we consider a general inhomogeneous situation, following the notation 
for classical fluids in Ref.~\cite{hans06}). Expressed in terms of center-of-mass 
and relative coordinates $X$ and $x_{12}$, the integral in~\eqref{eq:Hdd} is well defined 
despite the short-distance singularity $1/r_{12}^3$ provided $\rho^{(2)}(X,x_{12})$ is regular in the 
limit $x_{12}\to 0$ (for a mathematically precise statement see e.g. Ref.~\cite{tria19}). The disappearance 
of the singularity due to the vanishing angular average of $P_2(\hat{z}\cdot\hat{x}_{12})$ is spoiled, however, 
if the effective range of $V_{sr}(r_{12})$ is set to zero. Indeed, provided
the actual interaction becomes isotropic at short distances, the two-particle density is asymptotically 
determined by the one obtained from a pseudopotential approach where  
\begin{equation}
\rho^{(2)}(X, x_{12})\; \xrightarrow[x_{12}\to 0]\; \frac{\mathcal{C}_2(X)}{(4\pi)^2} \left( \frac{1}{r_{12}^2}- \frac{2}{ar_{12}}+\ldots\right)  
 \label{eq:contact}
 \end{equation} 
is singular at short distances~\cite{tan08energy,braa11}. Here, $\mathcal{C}_2(X)\to [4\pi\rho^{(1)}(X) a]^2$ is 
the Tan contact density in the weak interaction limit, which replaces 
the ill-defined local pair distribution function $g^{(2)}(0)$ for zero-range interactions (see Ref.~\cite{zwer21} for an introduction
to the Tan relations and the associated singular behavior of short distance correlations for Bosons). In practice, this
problem - which does not show up at the mean-field  level where $g^{(2)}(x_1,x_2)\to 1$ is replaced by one - is 
avoided if the expectation value of the dipolar interaction energy is evaluated with a two-particle density $\rho^{(2)}(X,x_{12})$
which is averaged over separations large compared to the actual interaction range. In particular, for $r_{12}\gg a$, the 
most singular term in Eq.~\eqref{eq:contact} is negligible and the next-to-leading term only gives rise to a small correction 
$\delta \rho^{(2)}(X, x_{12})= - [\rho^{(1)}(X)]^2\, 2 a/r_{12}$ to the mean-field result. As will be shown below, a cutoff at short
distances is also required to properly account for the angular dependence of the pair distribution function.\\

Beyond the issue of the singularity of the dipolar interaction at short distances for Bosons,
which generically have a finite probability to be at the same point in space, a major problem that needs to be 
addressed in dipolar fluids concerns the long-range and anisotropic, partially attractive nature of the interaction. 
Indeed, the standard proofs by Ruelle~\cite{ruel63} and Fisher~\cite{fish64stability} on the existence of a proper 
thermodynamic limit require interactions which decay faster than $1/r^3$ for large distances, leaving dipolar 
fluids as a marginal case. This problem has been discussed first by Griffiths~\cite{grif68} for dipoles on a lattice 
and then, more generally, also for continuum fluids in Refs.~\cite{froe78,bane98}. What has been shown by these 
authors is that
- in the presence of a stabilizing short distance regularization  e.g. via a lattice or a hard-sphere potential - the long-range 
and partially attractive dipolar interaction gives rise to an extensive free energy $F(T, V, N)=N\cdot \tilde{f}(T, v=V/N)$ 
with a free energy per particle $\tilde{f}$ which is finite and independent of the boundary conditions provided
the expectation value  $\mu_m\sum_j\langle\hat{d}_j\rangle\equiv 0$ of the total dipole moment vanishes. The proof thus 
covers standard polar fluids like water whose electric dipoles point in an arbitrary direction or magnetic systems in the absence of an
ordering external field. The situation is fundamentally different, however, for dipolar gases where the dipole directions $\hat{d}_j$ are 
all identical and their sum $\sum_j\hat{d}_j=N\,\hat{z}$ is extensive. As a consequence, even if the issue of the existence 
of a well defined limit $N\to\infty$ is ignored for typical particle numbers $N\simeq 10^4$~\cite{chom23},  standard 
results like the Gibbs-Duhem relation $\mu=\tilde{f}+pv$ are not expected to hold and, moreover, 
thermodynamic quantities will become shape-dependent. These points will be addressed in some detail in the following. \\

\subsection{Magneto-chemical potential and demagnetization tensor}
A well known feature in the thermodynamics of particles with charge $q$ 
is that the full chemical potential $\mu(x)+q\phi(x)$ whose gradient determines  
the current in an inhomogenous situation contains the interaction energy with the local electrostatic 
potential $\phi(x)$ in addition to the contribution $\mu(x)$ associated with other 
interactions.  In a completely analogous manner, for particles with 
permanent magnetic dipole moments pointing along $\hat{z}$, the relevant 
magneto-chemical potential contains a contribution $\mu_mB_z(x)$, where $B_z(x)$ 
is the exact magnetic field at the position of the particle. In a formal manner, this can 
be derived by noting that the local chemical potential 
\begin{equation}
\mu(x)=\frac{\delta F[\rho^{(1)}]}{\delta\rho^{(1)}(x)}=\mu_{\rm sr}(x)-\mu_mB_z(x)
\hspace*{0.2cm} {\rm with} \hspace*{0.2cm} -\mu_mB_z(x)=\int_{x'} V_{\rm dd}(x-x')\,\rho^{(1)}(x') g^{(2)}(x,x')
 \label{eq:mux}
 \end{equation} 
is obtained by a functional derivative of the free energy (or simply $\langle \hat{H} \rangle$ at $T=0$)
with respect to the one-particle density~\cite{hans06}. Here, the explicit expression for the exact local 
magnetic field due to all other dipoles follows from differentiation of the associated interaction energy~\eqref{eq:Hdd} while the 
contribution $\mu_{\rm sr}(x)$ arises from the short-range interactions, including the change in kinetic energy due to $\hat{H}_{\rm dd}$.
As a consequence of the long-range nature of the dipolar interaction, $\mu(x)$ is not - as usual - fully determined by temperature and 
the local one-particle density $\rho^{(1)}(x)$ but depends on the overall shape of the sample. This is evident 
already at the mean-field level $ g^{(2)}_{\rm mf}\equiv 1$ where the calculation of $B_z(x)$ reduces to a 
problem in classical magnetostatics with given magnetization $\mathbf{M}(x)=\mu_m\rho^{(1)}(x)\hat{z}$. 
The appearance of shape-dependent thermodynamics is well known in this context and it is
a standard textbook problem to show e.g. that the internal magnetic field in a sphere with constant 
magnetization is $\mathbf{B}_{\rm in}=(2/3)\mathbf{M}$~\cite{jack99}.  An exact solution is available 
also for more general geometries provided the magnetization current density $\mathbf{j}_M(x)={\rm rot}\;\mathbf{M}(x)$ 
may be approximated by a pure surface current. This applies naturally in solid ferromagnets with a sharp boundary and it carries 
over to the inner region of trapped or self-bound dipolar gases where the spatial dependence of $\mathbf{M}(x)$
is negligible.  The relevant demagnetization field $\mathbf{H}=-\underline{\mathbf{N}}\,\mathbf{M}$
in $\mathbf{B}=\mathbf{H}+\mathbf{M}$ is then determined by a shape-dependent demagnetization 
tensor $\underline{\mathbf{N}}$ which obeys ${\rm Tr}\;\underline{\mathbf{N}}=1$~\cite{land84}. Specifically, 
for a spheroid with aspect ratio $\kappa=R_x/R_z$ and $\mathbf{M}$ pointing along the $z$-axis, 
the tensor $\underline{\mathbf{N}}$ is diagonal. Symmetry fixes the relevant eigenvalue $n^{(z)}$ to be 
equal to $1/3$ in the case of a sphere while $n^{(z)}$ approaches zero for a long cylindrical spheroid 
with $\kappa\to 0$, where the demagnetization effect is negligible. Now, there is a subtle point 
which needs to be taken into account in an application of these results to the interaction induced 
field $B_z(x)$ that enters the local chemical potential~\eqref{eq:mux}.    
This has to do with the fact that the contribution $(2/3)\mathbf{M}$ to the inner field of a spherical
configuration comes entirely from the singular self-interaction term due to an effective delta 
function in the magnetic field of a point dipole~\cite{jack99}. In contrast to the corresponding zero-range contribution 
to the interaction between an electron and a nuclear magnetic moment which is the origin of hyperfine 
shifts in s-states~\cite{grif82}, it is physically reasonable to assume that there is no direct overlap between 
the partially filled electronic shells in rare-earth atoms. The effective delta function contribution must therefore 
be subtracted, which leads to $\mathbf{B}_{\rm dd}=\mathbf{H}+\mathbf{M}/3$ for the inner field due to the dipolar interaction. 
Using $\mu_m^2=3g_{\rm dd}$, the resulting contribution to the chemical potential at the mean-field level
\begin{equation}
\mu_mB_z^{\rm (mf)}(x)=g_{\rm dd}\left(1-3\,n^{(z)}\right)\rho^{(1)}(x) = g_{\rm dd}\, f(\kappa)\rho^{(1)}(x)
 \label{eq:mf}
 \end{equation} 
displays a simple shape-dependence via the demagnetization tensor element $n^{(z)}$.  
The explicit result for $n^{(z)}$ is derived e.g. in Ref.~\cite{land84} in the context of the equivalent 
problem of depolarization fields of perfect conductors. For an oblate spheroid with 
excentricity $e=\sqrt{\kappa^2-1}$ it is given by
\begin{equation}
n^{(z)}(e)=(1+e^2)\frac{e-\arctan{(e)}}{e^3}\;\to\; (1-\tilde{e}^2)\frac{\operatorname{artanh}{(\tilde{e})}-\tilde{e}}{\tilde{e}^3} = n^{(z)}(\tilde{e})\,\in(0,1/3]
 \label{eq:demag}
 \end{equation} 
which also determines the demagnetization factor $n^{(z)}(\tilde{e})$ in a prolate situation with excentricity $\tilde{e}=\sqrt{1-\kappa^2}$ 
by straightforward analytic continuation $e\to i\tilde{e}$. The expression~\eqref{eq:mf} reproduces
a result that has been derived early on in the field, based on a Gaussian Ansatz for the density profile
of a dipolar gas in a harmonic trap~\cite{yi01,giov03}.
It has been shown to hold also in the exact solution of the Gross-Pitaevskii equation in the 
Thomas-Fermi limit~\cite{eber05} but - somewhat surprisingly - the simple connection $f(\kappa)=1-3n^{(z)}$ of 
the shape function  $f(\kappa)$ with the standard demagnetization tensor of magnetostatics does not 
seem to have been realized in the literature so far. Note that due to $n^{(z)}(\tilde{e})<1/3$ in a prolate 
configuration $\kappa<1$,  the internal magnetic field $B_z(x)$ in the bulk of the atomic cloud is pointing 
along the positive $z$-direction. This leads to a lowering of the chemical potential with respect to 
the value $\mu_{sr}(x)$ determined by the short-range interactions, reflecting the dominance 
of attractively interacting dipoles in such a configuration. \\

An obvious question is to which extent corrections beyond mean-field which are contained in the deviation 
of the pair distribution function $g^{(2)}(x,x')$ in Eq.~\eqref{eq:mux} from the trivial limit one, do affect the 
result~\eqref{eq:mf}. It turns out that there is indeed an important qualitative change which arises from the 
anisotropy of the dipolar interaction. For a rotation invariant $V_{sr}(r_{12})$, the two-particle density $\rho^{(2)}(X, x_{12})$ 
only depends on the magnitude $r_{12}=|x_{12}|$ of the separation vector. This is no longer true for anisotropic 
interactions, where also the orientation of $x_{12}$ with respect to the dipole direction enters. The associated 
physics has been described in quantitative terms in a landmark paper by Wertheim on a classical fluid of hard 
spheres with point dipoles whose directions $\hat{d}_j$ fluctuate statistically~\cite{wert71}. Following his notation
\footnote{In Ref.~\cite{wert71}, an additional contribution $h_{\Delta}(r_{12})\hat{d}_1\cdot\hat{d}_2$ appears 
which couples to the relative orientation of the dipoles. In the present case with $\hat{d}_1\cdot\hat{d}_2\equiv 1$,
this contribution can be incorporated into the rotation invariant function $h_S(r_{12})$.},        
the pair correlation function of a homogeneous dipolar fluid with rotation invariance around the $z$-axis 
can be decomposed in the form
\begin{equation}
h^{(2)}(x_{12})=g^{(2)}(x_{12})-1=h_S(r_{12})+h_{D}(r_{12})\cdot 2P_2(\hat{z}\cdot\hat{x}_{12})+\ldots
 \label{eq:h12}
 \end{equation} 
Here, the dots indicate terms involving Legendre polynomials of higher order which are neglected since
their contribution to $\mu_mB_x(x)$ contains angular integrals $\int_{\Omega} P_2\,P_l$ which tend to cancel 
(only even $l$ appear as long as $z\to -z$ is a symmetry). The two scalar functions $h_S(r_{12})$ and $h_D(r_{12})$ describe 
quite different physics and turn out to have opposite sign. Specifically, $h_S(r_{12})<0$ accounts for 
an effective short-range repulsion, reducing the probability to find two particles 
at separation $r_{12}$ below that of an uncorrelated state. In turn, at fixed $r_{12}$, a positive $h_D(r_{12})>0$ 
favors attractive head-to-tail configurations with $(\hat{z}\cdot\hat{x}_{12})^2>1/3$.
In explicit form, the different behavior of separation and angular orientation correlations with respect to the $z$-axis 
may be derived for a uniform Bose gas in the limit of weak dipolar interactions $\varepsilon_{\rm dd}=a_{\rm dd}/a\ll 1$ by 
expanding the static structure factor within the Bogoliubov approximation in the regime of small momenta  
\begin{equation}
S(q)=1+nh^{(2)}(q)\; \xrightarrow[\rm Bog.]\; \left[1+2nV(q)/\varepsilon_q\right]^{-1/2}\; \xrightarrow[|q|\xi, \varepsilon_{\rm dd}\ll 1]\; 
\frac{|q|\xi}{\sqrt{2}}\left[ 1- \varepsilon_{\rm dd} P_2(\hat{z}\cdot\hat{q})+\ldots\right]\, .
 \label{eq:Sbog}
 \end{equation} 
Here, $V(q)=g+V_{\rm dd}(q)$ is the Fourier transform of the interaction, $\xi=1/\sqrt{8\pi na}$ the healing length 
associated with its short-range part and $\varepsilon_q=\hbar^2q^2/(2m)$ the free particle dispersion. The 
expansion~\eqref{eq:Sbog} of the static structure factor at low momenta precisely matches the corresponding 
one in real space of Eq.~\eqref{eq:h12}. Moreover, the non-analytic behavior in $q$ implies that the 
associated pair correlation functions $nh_S(r)\to -\xi/(\sqrt{2}\pi^2 r^4)$ and $nh_D(r)\to +\varepsilon_{\rm dd}\cdot\sqrt{2}\xi/(\pi^2 r^4)$ 
follow a power law decay at $T=0$ and have opposite sign. A further point which distinguishes the correlations of the magnitude of separation vectors 
from those which describe their orientation with respect to the dipole direction shows up at finite temperature: while $h_S(r)$ 
decays to zero exponentially beyond the thermal wavelength $\lambda_T$, the angle dependent contribution 
\begin{equation}
h_D(r)\,\big|_{T\ne 0}\to h_{\infty}/r^3 \hspace*{0.2cm} {\rm with} \hspace*{0.2cm} nh_{\infty}=\frac{3}{4\pi}\frac{(\varepsilon(0)-1)^2}{\varepsilon(0)}\frac{k_BT}{g_{\rm dd}n}
 \label{eq:hD}
 \end{equation} 
retains a power law decay, following that of the dipolar interaction. The asymptotic behavior~\eqref{eq:hD} is an exact result first derived
by Wertheim~\cite{wert71} for classical fluids with randomly oriented electric dipoles and $\varepsilon(0)$ the associated static dielectric
constant.  It is consistent with a correlation inequality at  finite temperature which implies that correlations cannot decay faster than the interactions. 
Using the connection $S(q)\to k_BT/mc^2(\hat{q})$ between the static structure factor at low momenta and the angle dependent 
sound velocity $c(\hat{q})$ of dipolar gases~\cite{bism12}, it is straightforward to show that - to linear order in $\varepsilon(0)-1\to \varepsilon_{\rm dd}\ll 1$,
where the homogeneous system with average density $n$ is stable - the result~\eqref{eq:hD} continues to hold for quantum fluids of magnetic 
dipoles which are fully polarized.\\

An important consequence of the results above for the thermodynamics of dipolar gases shows up by considering 
nearly spherical configurations, where mean-field predicts that the dipolar interactions average to zero. Obviously, 
this result remains valid more generally for an arbitrary pair correlation function which is rotation invariant. Including 
the angular dependent contribution $h_D(r)\ne 0$, however, there is a non-vanishing magnetic field 
\begin{equation}
\mu_mB_z(x)\big|_{\kappa=1}=(12/5) g_{\rm dd}\,\rho^{(1)}(x)\cdot K \hspace*{0.3cm} {\rm with} \hspace*{0.3cm} K=\int_0^{\infty} dr\frac{h_D(r)}{r}>0
 \label{eq:K}
 \end{equation} 
near the center which points along the positive $z$-direction. The dimensionless parameter $K$, which plays a central role
in a microscopic theory of the dielectric constant in classical polar fluids
\footnote{Combining point dipoles with hard spheres, the dielectric constant is a function only of the volume fraction $\eta=(\pi/6)n\sigma^3$
and the ratio $y=g_{\rm dd}n/(3k_BT)$. Remarkably, this extremely oversimplified model reproduces quite well the observed value $\varepsilon(0)=78.4$ 
for water under normal conditions, where $\eta\simeq 0.42$ and $y\simeq 3.35$, see Fig. 11.5 in Ref.~\cite{hans06}.},
apparently requires a cutoff at short distances. As mentioned above, a description of dipolar gases beyond mean-field must
therefore take into account the finite range of interactions which, in practice, is of the order of the short-range scattering 
length $a$. An experimental estimate of the parameter $K$ may be obtained by observing the finite critical strength
$\varepsilon_{\rm dd}^{(c)}(\kappa=1)=5/(12\,K)$ of the dipolar interaction where even a spherical cloud becomes unstable 
because the chemical potential $\mu(x)\approx g\rho^{(1)}(x)-\mu_mB_z(x)$ turns negative at leading order.

\subsection{Pressure and uniaxial tension from dilatations}
In the discussion so far, the consequences of the anisotropy and long-range nature of 
the dipolar interctions have been addressed by assuming a given shape, characterized 
e.g. by the aspect ratio $\kappa$. In reality, the specific shape of the atomic cloud is not an independent
variable but is determined by the interaction strength and the external confinement, if present.     
In thermal equilibrium, it follows from the balance of forces which - in quite general form - may 
be expressed in terms of the associated momentum current tensor $\underline{\Pi}(x)$ by
\begin{equation}
\nabla\,\underline{\Pi}(x) \; \xrightarrow[\rm equ.]\; 
\left( \begin{array}{c} \partial_x \Pi^{xx}\\  \partial_y \Pi^{yy}\\  \partial_z \Pi^{zz} \end{array} \right)
= \left\{ \begin{array}{cc} 0  &  \hspace*{0.4cm} \mbox{self-bound} \\  
-\rho^{(1)}_{\phi}(x)\,{\rm grad}\,\phi(x) &  \mbox{trapped}
\end{array} \right.
\label{eq:hydro}
\end{equation}
Here, $\phi(x)$ is the external trap potential and $\rho^{(1)}_{\phi}(x)$ the resulting equilibrium value
of the inhomogeneous one-particle density. Moreover, we have used the 
fact that in thermal equilibrium the tensor $\underline{\Pi}(x)$ has no off-diagonal elements
as would appear e.g. in the presence of a finite Hall viscosity. Now, for rotation 
invariant interactions, all diagonal elements of $\underline{\Pi}(x)$ are equal and they define a local 
pressure $p(x)$. The left hand side of Eq.~\eqref{eq:hydro} is then just the gradient of pressure  
in a standard hydrostatic equilibrium. For a harmonic confinement, where 
the external contribution to the total Hamiltonian 
\begin{equation}
\hat{H}_{\omega}=\hat{H}+\frac{1}{2} {\rm Tr}\,\left(\underline{\omega}^2\,\underline{\hat{\theta}}\right) 
\hspace*{0.3cm} {\rm with} \hspace*{0.3cm} \left(\underline{\hat{\theta}}\right)^{ij}=\int_x x^ix^j\hat{\rho}(x)
\label{eq:harmonic}
\end{equation}
can be expressed in terms of a tensor $\underline{\omega}^2$ of the trap frequencies and the 
inertia tensor $\underline{\hat{\theta}}$ defined via the mass density operator $\hat{\rho}(x)$,  
the resulting cloud shape follows from the exact relation 
\begin{equation}
\langle \underline{\hat{\theta}}\rangle_{\text{eq}}= (\underline{\omega}^2)^{-1}\,\int_x p(x) = E_{\text{rel}}\, (\underline{\omega}^2)^{-1}
\hspace*{0.3cm} {\rm for\; rotation\; invariant} \hspace*{0.3cm} V(x_{12})=V(r_{12})
\label{eq:inertia}
\end{equation}
obtained by taking the scalar product of Eq.~\eqref{eq:hydro} with $\mathbf{x}$ and integrating over all space. 
Note that the relation~\eqref{eq:inertia} provides only an implicit result for the cloud shape as characterized by its inertia tensor
since the release energy $E_{\text{rel}}=\int_x p(x)$ is itself shape-dependent. An exact consequence of Eq.~\eqref{eq:inertia}, however,
is that the aspect ratio which - irrespective of the detailed density profile - is defined by $\kappa=[\theta^{xx}/\theta^{zz}]^{1/2}$
coincides with the ratio $\lambda=\omega_z/\omega_x$ of the trap frequencies (we assume $\omega_x=\omega_y\ne\omega_z$
in general, i.e. there is rotation invariance around the $z$-axis). \\

The anisotropy of the dipolar interaction invalidates the simple relation~\eqref{eq:inertia} and leads to 
shapes  with $\kappa<\lambda$ which are elongated along the $z$-direction since the energy is lowered 
if the dipoles tend to arrange in a head-to-tail configuration. Dipolar gases are thus uniaxial quantum 
fluids which  - in quite  general terms - may be defined by $\Pi^{xx}=\Pi^{yy}\ne\Pi^{zz}$ in equilibrium.  
To characterize their size and shape one needs two instead of just a single scalar function. Specifically, 
we will use ${\rm Tr}\;\underline{\Pi}$ and the anisotropy $h=\Pi^{xx}-\Pi^{zz}$, which acts like a uniaxial tension tending 
to stretch the system along $z$. As will be discussed below, the system length $L_z$ will then appear 
as an additional thermodynamic variable which is conjugate to the field $h$. On a microscopic level, an explicit expression for
both ${\rm Tr}\;\underline{\Pi}(x)$ and $h(x)$ may be derived by extending the results of Martin and Schwinger~\cite{mart59} 
for the momentum current tensor in a non-relativistic quantum many-body problem with rotation invariant interactions.  
Quite generally, pressure - as the field conjugate to a local expansion - is defined 
by the change in energy density under a uniform isotropic dilatation $x\to bx$ with scale factor $b\to 1$. 
This only fixes ${\rm Tr}\;\underline{\Pi}(x)$, however. In order to specify the two independent 
components  $\Pi^{xx}$ and $\Pi^{zz}$ it is necessary to consider, in addition,
the volume preserving dilatation $x\to \theta[x]$ where the components $z\to\theta z$ and $ (x,y) \to (x,y)/\sqrt{\theta}$ 
are scaled differently. 
The full momentum current tensor of uniaxial quantum fluids is thus obtained from the 
microscopic Hamiltonian density $\hat{\mathcal{H}}(x)$ in $\hat{H}=\int_x \hat{\mathcal{H}}(x)$ by 
evaluating the response to the two independent dilatations (note the different signs)
\begin{equation}
{\rm Tr}\,\underline{\Pi}(x)=2\Pi^{xx}(x)+\Pi^{zz}(x)=-\frac{\partial \langle\hat{\mathcal{H}}(bx)\rangle_{\rm eq}}{\partial b} \,\Big|_{b=1}
\hspace*{0.3cm} {\rm and} \hspace*{0.3cm} h(x)= \frac{\partial \langle\hat{\mathcal{H}}(\theta[x])\rangle_{\rm eq}}{\partial \theta}  \,\Big|_{\theta=1}\, .
\label{eq:dila}
\end{equation}
In explicit form, the anomalous field $h(x)$ receives contributions only from the kinetic energy and the dipolar interaction. 
They can be expressed in terms of the field operators $\hat{\psi}$ as 
\begin{equation}
h(x)= \frac{\hbar^2}{m}\left[ \frac{1}{2} \langle \nabla{\!}_{\perp}\hat{\psi}^{\dagger}\cdot\nabla{\!}_{\perp}\hat{\psi}\rangle_{\rm eq}(x) -  
\langle \partial_z\hat{\psi}^{\dagger}\partial_z\hat{\psi}\rangle_{\rm eq}(x)\right] +\partial_{\theta} \mathcal{D}(\theta[x]) \,\big|_{\theta=1}\, .
\label{eq:hx}
\end{equation}
Here, $\nabla{\!}_{\perp}$ denotes the gradient with respect to the radial coordinates $(x,y)$ and $ \mathcal{D}(x)$ 
is a dipolar analog of the contact density $\mathcal{C}_2(x)$ for zero-range interactions. It is defined by the expectation 
value of the dipolar interaction energy density  
\begin{equation}
\mathcal{D}(x) = \langle\hat{\mathcal{H}}_{\rm dd}(x)\rangle_{\rm eq}= -\frac{1}{2}\langle\hat{M}_z(x)\hat{B}_z(x)\rangle_{\rm eq}
\label{eq:Dx}
\end{equation}
which involves the product of the operators for the magnetization $\hat{M}_z(x)$ and the 
field $\hat{B}_z(x)$ from all other dipoles at position $x$. Due to the long-range nature of 
the interaction, $\mathcal{D}(x)$ is in fact non-local and shape-dependent, a point which 
will be discussed further below. This is very different from the energy density $\varepsilon_{\rm sr}(x)$ 
and the associated contribution~\cite{wern12,zwer21}
\begin{equation}
p_{\rm sr}(x)=\frac{2}{3}\varepsilon_{\rm sr}(x)+\frac{\hbar^2}{24\pi m a}\mathcal{C}_2(x)\; \xrightarrow[\rm mf]\; \frac{g}{2}\,[\rho^{(1)}(x)]^2 +\ldots
\label{eq:psr}
\end{equation}
to the pressure arising from short-range interactions which - at the mean-field level - is quadratic in the local 
one-particle density $\rho^{(1)}(x)$ and independent of the shape. Using the microscopic definition~\eqref{eq:dila},
the full momentum current tensor of dipolar gases in turn contains the shape-dependent contributions $\mathcal{D}(x)$ 
and $\partial_{\theta} \mathcal{D}(\theta[x]) \,\big|_{\theta=1}$ according to 
\begin{equation}
\Pi^{xx}(x)=p_{\rm sr}(x)+\mathcal{D}(x)+\frac{1}{3}\,h(x)\, .
\label{eq:Pixx}
\end{equation}

\vspace*{0.2cm}
The formal relations above turn out to lead to observable and intuitively plausible consequences 
for the density profile of harmonically trapped dipolar gases. In a situation with $\omega_x=\omega_y$, where 
one has rotation invariance around the $z$-axis, a radial integration of the $(x,y)$-components of the 
equilibrium condition~\eqref{eq:hydro} shows that the integrated column density 
\begin{equation}
\tilde{n}_{\phi}(z)=\int\! dxdy\,\rho^{(1)}_{\phi}(x,y,z)= \frac{2\pi}{m\omega_x^2}\,\Pi^{xx}(0,0,z)
\label{eq:colum}
\end{equation}
at a given $z$ is a direct measure of the $xx$-component of the momentum current tensor along
the center of the atomic cloud. This relation is equivalent to one derived by Ho and Zhou for rotation invariant 
interactions in the context of imbalanced Fermi gases~\cite{ho10}. A different result appears, however, for
the $z$-derivative of the density profile which follows from the third component of the momentum balance~\eqref{eq:hydro}.
The fact that $\Pi^{xx}$ and $\Pi^{zz}$ no longer coincide in the uniaxial case leads to a violation of the Gibbs-Duhem
relation which shows up as a non-vanishing contribution on the right-hand side of the equation
\begin{equation}
\frac{d\tilde{n}_{\phi}(z)}{dz}+2\pi\frac{\omega_z^2}{\omega_x^2} z\,\rho^{(1)}_{\phi}(0,0,z) = \frac{2\pi}{m\omega_x^2}\,\partial_z h(0,0,z)> 0\, .
\label{eq:GD}
\end{equation}
Since the anomalous field $h(x)$ is negative and decays along the $z$-direction away from the center, the profile $\tilde{n}_{\phi}(z)$ 
falls off more slowly than what is expected for rotation invariant interactions. The uniaxial nature of dipolar gases
and the associated field $h(x)$ can thus be inferred directly from in-situ density profiles.  

\subsection{Thermodynamic relations and the tensor virial theorem} 
The appearance of a finite uniaxial tension in dipolar gases leads to 
a fundamental change in thermodynamic relations. This turns out to
provide an understanding of the shape-dependence mentioned above
and also the peculiar form of self-bound droplets that will be discussed
in the following subsection. In order to derive the extension to uniaxial fluids of the 
standard relation $dF=-pdV$ for the change in free energy associated with a shape 
deformation at a fixed value of temperature and particle number it is 
convenient to introduce a local deformation tensor $\underline{u}(x)$ 
as in solids. Quite generally, $\underline{u}(x)$ is the conjugate field to the momentum 
current (or stress) tensor $\underline{\Pi}(x)$~\cite{chai95} which, by the definition of an 
uniaxial fluid in thermal equilibrium, is diagonal with just two independent components. 
As a consequence, the associated differential of the free energy  
\begin{equation}
dF[{\rm Tr}\,\underline{u}, u_{zz}] \big|_{T,N} =-\int_x {\rm Tr}\, \bigl[\underline{\Pi}(x)d\underline{u}(x)\bigr]
\; \xrightarrow[\rm fluid]\; \int_x \left[ -\Pi^{xx}(x)\, d\,{\rm Tr}\,\underline{u}(x) + h(x)\,du_{zz}(x)\right]
\label{eq:ela}
\end{equation}
depends on the two dimensionless variables ${\rm Tr}\,\underline{u}$ and $u_{zz}$. In physical terms,
they describe a local change $d\,{\rm Tr}\,\underline{u}(x)=dv(x)/v$ in volume per particle at a fixed 
extension along the $z$-direction or a change of length $d\,u_{zz}=dL_z/L_z$ at fixed volume
\footnote{While a cloud of atoms has no sharp boundary beyond the Thomas-Fermi limit where
the kinetic energy is negligible, the associated Thomas-Fermi radius $R_z$ still provides an appropriate 
measure for the length $L_z\to 2R_z$.}. The relation~\eqref{eq:ela} shows that for a uniaxial fluid $\Pi^{xx}\to p$ 
can be identified with the pressure as the variable conjugate to volume at fixed length. In addition, the 
anomalous field $h(x)$ gives rise to a line tension contribution in the Gibbs relation     
\begin{equation}
dF(T,V,L_z,N)=-SdT-pdV+h_1dL_z+\mu dN \hspace*{0.3cm} {\rm with} \hspace*{0.3cm} h_1L_z=\int_x h(x)\, ,
\label{eq:Gibbs}
\end{equation}
favoring an increase of length $L_z$ at given volume to lower the free energy ($h_1<0$).
As will be shown below, the fact that $L_z$ appears as an additional extensive thermodynamic 
variable leads to a violation of the Gibbs-Duhem relation. Typically, such violations become negligible 
in the thermodynamic limit, which holds e.g. for a finite surface tension $\sigma$ in a self-bound liquid 
state described by $dF=\sigma\, dA$ at fixed $T, N, V$. For short-range interactions, this
contribution scales with the surface area $A\sim N^{2/3}$ and thus eventually becomes subdominant 
compared with the extensive terms. The situation is different here, however, due to the long-range nature
of the dipolar interaction. Indeed, since the interaction contribution $\partial_{\theta} \mathcal{D}(\theta[x]) \,\big|_{\theta=1}$ in 
Eq.~\eqref{eq:hx} to the underlying variable $h(x)$ is intensive, $dF=h_1\, dL_z\sim N$ remains extensive 
for an arbitrary large system. In physical terms, this can be understood
from the fact that the dipolar contact density $\mathcal{D}(x)$ in Eq.~\eqref{eq:Dx} arises from the    
magnetic field which, as discussed in subsection 2.1, is effectively generated by surface currents. 
The demagnetization field $\mathbf{H}(x)=-{\rm grad}\,\phi_M(x)$ is thus non-local, arising
from magnetic surface 'charges' with density $\sigma_M=\mathbf{n}\cdot\mathbf{M}$  ($\mathbf{n}$ is the unit normal vector)~\cite{jack99}.
As emphasized in Ref.~\cite{bane98}, the $1/r^2$ decay of the field is precisely cancelled by the increase $\sim r^2$
of the surface area.  Both the magnetic field and the associated energy density $\mathcal{D}(x) =  \langle\hat{\mathcal{H}}_{\rm dd}(x)\rangle_{\rm eq}$
are therefore independent of the system  size.   \\

The extended Gibbs relation~\eqref{eq:Gibbs} implies that - at fixed temperature $T$ - the pressure 
and chemical potential involve the strain tensor element $u_{zz}$ as a further thermodynamic variable
beyond the volume per particle $v$. The associated dependence is determined by the Maxwell relations 
\begin{equation}
v\frac{\partial p}{\partial v} = \frac{\partial \mu}{\partial v}  \hspace*{0.4cm} {\rm {at\; fixed}}\, u_{zz}\, ,  \hspace*{0.8cm} 
\frac{\partial \mu}{\partial u_{zz}} = \frac{\partial h}{\partial (1/v)} \hspace*{0.8cm} {\rm and} \hspace*{1cm} \frac{\partial p}{\partial u_{zz}}=v\frac{\partial h}{\partial v}\, .
\label{eq:Maxwell}
\end{equation}
In an inhomogeneous situation, where the density $\rho^{(1)}(x)$ and the field $h(x)$ are spatially varying, the 
standard Gibbs-Duhem relation $d\mu\vert_T=vdp\vert_T$ is therefore violated. This leads to an anomalous contribution 
to the axial derivative of the density profile of a harmonically trapped dipolar gas derived in Eq.~\eqref{eq:GD}. 
A further consequence of the dependence of pressure $p(v,u_{zz},T)$ on the uniaxial strain $u_{zz}$ is the angular 
dependence of the sound velocity in a uniform dipolar gas.  In quite general terms, the difference $\Delta f=f-f_{\rm eq}$
in the free energy density with respect its value $f_{\rm eq}$ in a homogeneous situation with 
vanishing deformations $({\rm Tr}\, \underline{u})_{\rm eq}=(u_{zz})_{\rm eq}\equiv 0$ can be written in the form of 
an elastic free energy of a uniaxial solid~\cite{chai95}
\begin{equation}
\Delta f [{\rm Tr}\,\underline{u}, u_{zz}]=\frac{1}{2}\left( u_{xx}+u_{yy}\, , u_{zz} \right) \left( \begin{array}{cc} K_{11} & K_{13} \\
K_{13} & K_{33} \end{array} \right) \left( \begin{array}{c} u_{xx}+u_{yy} \\ u_{zz} \end{array} \right) = 
\frac{K_{11}}{2}\left[{\rm Tr}\,\underline{u}+\tilde{\gamma} u_{zz}\right]^2 +\frac{\tilde{B}}{2}u_{zz}^2
\label{eq:solid}
\end{equation}
which is rotation invariant in the $x,y$-plane. The condition that Eq.~\eqref{eq:solid} describes a fluid rather than a solid requires 
that the associated tensor of elastic constants has vanishing determinant $\tilde{B}K_{11}=K_{33}K_{11}-K_{13}^2=0$. 
The last term is then absent and there is only a single sound mode 
\footnote{For finite values of the parameter $\tilde{B}$, the free energy~\eqref{eq:solid} describes a smectic-A phase 
with broken translation invariance along the $z$-direction~\cite{chai95}. It exhibits two independent sound modes, both 
in the case of a classical fluid~\cite{mart72} and also - in the quite different form of a wave-like propagation of defects 
in supersolids predicted by Andreev and Lifshitz~\cite{andr69} - in the superfluid~\cite{hofm21}. 
These two modes have very recently been observed in a 2D smectic phase of a Bose Einstein condensate 
in the presence of a periodic in time modulation of the scattering length~\cite{lieb24}, allowing to determine both the superfluid 
fraction $f_s$ and the effective layer compression modulus $\tilde{B}$ in quantitative terms.}.
Its velocity depends on the direction with respect to the orientation of the dipoles according 
to $\bar{\rho}c^2(\hat{q})=K_{11}+(K_{33}-K_{11})\cdot (\hat{z}\cdot\hat{q})^2$~\cite{chai95,mart72}.
This angular dependence is a result of the coupling between density and strain described by the 
dimensionless parameter $\tilde{\gamma}$ in
\begin{equation}
h=\frac{\partial f}{\partial u_{zz}}\, \,\Big|_{{\rm Tr}\,\underline{u}} \to \;
\tilde{\gamma}K_{11}\left({\rm Tr}\,\underline{u}+\tilde{\gamma} u_{zz}\right)  \hspace*{0.4cm} {\rm with} \hspace*{0.4cm} 
\tilde{\gamma}=\frac{K_{13}}{K_{11}}-1\; \xrightarrow[\rm mf]\; \left(\frac{1+2\varepsilon_{\rm dd}}{1-\varepsilon_{\rm dd}}\right)^{1/2}-1 \, .
\label{eq:tg}
\end{equation}
A physical understanding of this coupling is provided by the observation 
that - according to Eq.~\eqref{eq:solid} - the energy cost for a compression ${\rm Tr}\,\underline{u}<0$
of a uniform dipolar gas can be exactly compensated by an expansion along $z$ with $u_{zz}=-{\rm Tr}\,\underline{u}\,/\tilde{\gamma}>0$.    
The mean-field result for $\tilde{\gamma}$ in terms of the dimensionless strength $\varepsilon_{\rm dd}$ 
of the dipolar interaction follows from the angular dependence $V(q)=g[1+\varepsilon_{\rm dd}\cdot 2P_2(\hat{z}\cdot\hat{q})]$
of the two-body interaction. It indicates that the uniform gas with a homogeneous mass density $\bar{\rho}$      
is stable only as long as $\varepsilon_{\rm dd}<1$, a point which will be discussed in detail in the following subsection.\\

An exact relation that holds for many-body problems in both classical and 
quantum physics is the virial theorem. It is a scalar identity which follows by 
considering the change in energy under a uniform dilatation of the coordinates 
$(x_1\ldots x_N)\to (b x_1\ldots bx_N)$~\cite{wern08}. Now, as discussed in Eq.~\eqref{eq:dila}
above, uniform dilatations are not sufficient to characterize the thermodynamically relevant set 
of shape deformations in the case of interactions which are not invariant under rotation. To deal 
with this more general situation, it is necessary to extend the virial theorem to a tensorial form, 
as was first derived by Parker~\cite{park54} for a classical system 
of point particles in an astrophysical context. The extension to the quantum many-body problem 
is straightforward in principle. Indeed, the virial theorem in its most general form simply 
states that the expectation value of the commutator
\begin{equation}
\langle i[\hat{H}+\hat{H}_{\rm ex}, \underline{\hat{D}}]\rangle \equiv 0 \; \xrightarrow[\rm harm.]\; 
\langle i[\hat{H}, \underline{\hat{D}}]\rangle=\langle\underline{\omega}^2 \underline{\hat{\theta}}\rangle 
\hspace*{0.4cm} {\rm with} \hspace*{0.4cm} (\underline{\hat{D}})^{ij}=\int_x x^i\hat{g}^j(x)/\hbar
\label{eq:virial}
\end{equation}
of the total Hamiltonian with the dilatation tensor operator $\underline{\hat{D}}$ vanishes in eigenstates of $\hat{H}+\hat{H}_{\rm ex}$. 
The dimensionless dilatation tensor involves the components $\hat{g}^j(x)$ of the momentum density operator and 
generates dilatations along the $j$-direction proportional to separate components $x^i$ of the coordinates~\cite{nish07}.   
In the particular case of a harmonic confinement, the contribution of $\hat{H}_{\rm ex}$ is 
reduced to the expectation value of the tensor contraction $\underline{\omega}^2 \underline{\hat{\theta}}$. Its trace
yields twice the trap energy $E_{\rm tr}$ according to Eq.~\eqref{eq:harmonic}. For rotation invariant interactions,
the equilibrium expectation value $\langle i[\hat{H}, \underline{\hat{D}}]\rangle_{\rm eq}=\underline{1}\cdot\int_x p(x)$
is proportional to the unit tensor.  The tensorial form of the virial theorem then reduces 
to the standard one obtained from a uniform dilatation $x\to bx$. 
In particular, the moment of inertia tensor is given by Eq.~\eqref{eq:inertia}. This is violated in the uniaxial situation, where 
the aspect ratio $\kappa=[\theta^{xx}/\theta^{zz}]^{1/2}$ no longer coincides with the ratio $\lambda=\omega_z/\omega_x$ of 
external trap frequencies. The difference is determined by the exact relation 
\begin{equation}
-\int_x h(x)=\omega_z^2\theta^{zz}-\omega_x^2\theta^{xx}=2E_{\rm tr}\cdot\frac{\lambda^2-\kappa^2}{\lambda^2+2\kappa^2} >0
\label{eq:kappa}
\end{equation}
which follows from the tensor virial theorem~\eqref{eq:virial} in an equilibrium state by considering the volume preserving dilatation $x\to\theta[x]$ 
introduced in Eq.~\eqref{eq:dila}. A detailed discussion of the connection between $\kappa$ and $\lambda$ as a function of
the dipolar interaction strength $\varepsilon_{\rm dd}$ has been given by Eberlein et al.~\cite{eber05} within the Thomas-Fermi approximation,
where the density profile $\rho^{(1)}_{\phi}(x)$ has a sharp boundary. In particular, they have shown that, in the presence of a strong 
external confinement $\lambda\gg 1$ along the $z$-direction, a dipolar gas remains stable even in the limit $\varepsilon_{\rm dd}\gg 1$ of 
negligible short-range repulsion. This has been verified experimentally by Koch et al.~\cite{koch08} for particle 
numbers $N\simeq 10^4$. Effects beyond mean-field, which are contained in Eq.~\eqref{eq:kappa}, 
therefore do not give rise to qualitative changes in a tightly confined situation. As will be shown below,
the situation is quite different in self-bound droplets whose form is completely determined by such corrections. 
A second independent relation which follows from the tensorial virial 
theorem is connected with the uniform dilatation $x\to bx$ generated by the scalar operator $\hat{D}={\rm Tr}\;\underline{\hat{D}}$. 
It can be written in terms of the local contact densities $\mathcal{C}_2(x)$ and $\mathcal{D}(x)$ associated 
with the short-range and dipolar interactions in the form
\begin{equation}
\langle i[\hat{H}, \hat{D}]\rangle=\langle\hat{H}\rangle+E_{\rm tr}+\frac{\hbar^2}{16\pi ma}\int_x \mathcal{C}_2(x) + \frac{1}{2}\int_x \mathcal{D}(x)= 2E_{\rm tr} \, .
\label{eq:virial-trap}
\end{equation}
A similar relation was derived in Ref.~\cite{hofm20} for dipolar gases in 2D, however there are two 
important differences:  since the dipolar interaction in 2D also leads to s-wave scattering, 
the two-body contact density $\mathcal{C}_2(x)$ is defined by the adiabatic derivative of the 
energy with respect to the full scattering length of the combined interaction $V_{\rm sr}+V_{\rm dd}$. 
As a consequence, the exact expression for the dipolar contact density $\mathcal{D}(x)$ requires
a subtraction in the pair distribution function at short distances associated with the 
contribution of  $V_{\rm sr}(r_{12})$~\cite{hofm20}. A second point is that the angular independent dipolar 
interactions $\sim \mu_m^2/r_{12}^3$ in 2D are effectively of a short-range nature. In contrast to the 
situation here, the associated contact density $\mathcal{D}(x)$ is then shape-independent.    

\subsection{Self-bound droplets and the LHY correction}
One of the major new developments associated with ultracold dipolar gases was the surprising observation of 
self-bound droplets by Ferrier-Barbut et al.~\cite{ferr16}. These droplets are very different from what 
is expected for droplets of a standard liquid, whose equilibrium configuration is spherical due to the presence of 
a finite surface tension $\sigma$. Indeed, as will be shown below, the combination of anisotropy and the 
dominantly attractive nature of dipolar interactions in a non-confined situation leads to droplets whose
aspect ratio $\kappa(N)\sim N^{-1/4}$ approaches zero for large particle numbers
\footnote{This scaling law has been seen originally in numerical simulations by Baillie et al.~\cite{bail17} 
and has recently been derived in analytical form by Dalibard~\cite{dali24}, based on a different but physically 
equivalent line of arguments.}.  \\

Based on the quite general hydrostatic equilibirium condition~\eqref{eq:hydro}, the existence of stable self-bound droplets  
requires that the divergence $\nabla\,\underline{\Pi}(x)\equiv 0$ of the momentum current tensor vanishes.
In a uniaxial fluid with long-range and angle-dependent attractive interactions, this condition is obeyed only if 
both ${\rm Tr}\;\underline{\Pi}(x)$ and $h(x)$ are zero. In contrast to the situation in stars, where a 
single scalar equation for the balance of the radial component of the internal and gravitational pressure is sufficient 
to determine the equilibrium shape (see e.g. Ref.~\cite{bali99stars} for an instructive discussion), 
two separate conditions need to be fulfilled here. In order to formulate these in a physically intuitive manner, 
it is convenient to rewrite the virial theorem~\eqref{eq:virial-trap} in a form where the external trap is eliminated  
\begin{equation}
\langle i[\hat{H}, \hat{D}]\rangle=\int_x \left[ 2\,\varepsilon_{\rm sr}(x)+\frac{\hbar^2}{8\pi ma}\mathcal{C}_2(x) + 3\,\mathcal{D}(x)\right]=
3\int_x \left[ p_{\rm sr}(x)+\mathcal{D}(x)\right]=\int_x {\rm Tr}\;\underline{\Pi}(x)
\label{eq:virial-D}
\end{equation}
Here, following the Tan pressure relation in Eq.~\eqref{eq:psr}, we have introduced the 
local pressure $p_{\rm sr}(x)$ which arises from the short-range interactions
\footnote{In the application of the Tan relations to Bosons, we consistently neglect three-body correlations which are 
connected with the Efimov effect and the underlying dependence of $ \langle\hat{\mathcal{H}}(x)\rangle_{\rm eq}$ on
the short-range parameter $\kappa_{*}$~\cite{braa11,wern12,zwer21}.}. 
Beyond the associated energy density $\varepsilon_{\rm sr}(x)$, it involves the two-body contact density $\mathcal{C}_2(x)$ 
which is defined by the adiabatic derivative of $ \langle\hat{\mathcal{H}}(x)\rangle_{\rm eq}$ with respect to the inverse scattering 
length $1/a$~\cite{braa11,wern12}.  Eq.~\eqref{eq:virial-D} shows that the condition of a vanishing ${\rm Tr}\;\underline{\Pi}(x)\equiv 0$ 
is equivalent to a local pressure balance $p_{\rm sr}(x)+\mathcal{D}(x)=0$, with $\mathcal{D}(x)$ playing the role of the pressure 
associated with the dipolar interactions. Prior to a detailed discussion of this exact and quite intuitive relation a number 
of comments should be added:
\begin{itemize}
\item The fact that the pressure arising from dipolar interactions is equal to the associated energy density $\mathcal{D}(x)$
is a consequence of the assumption that the interaction $V_{\rm dd}(x_{12})\sim \mu_m^2/r_{12}^3$ follows an inverse cube power law at all distances.\\

\item Despite the formal notation, $\mathcal{D}(x)$ is not a local variable but depends on the overall shape of the droplet, as becomes
evident in the mean-field result~\eqref{eq:hmf} below.\\

\item The separation of pressure into a short-range and a dipolar contribution follows from the additivity 
$V=V_{\rm sr}+V_{\rm dd}$ of the interactions in the underlying Hamiltonian. It does not imply that pressure is simply additive, 
however, which is true only in mean-field. In general, the contributions $p_{\rm sr}(x)$ and $\mathcal{D}(x)$ will 
depend on both $g$ and $g_{\rm dd}$. \\ 

\item The identification of ${\rm Tr}\;\underline{\Pi}(x)$ with $3 \,[p_{\rm sr}(x)+\mathcal{D}(x)]$ on a local
level neglects a possible divergence which integrates to zero in the virial equation. Indeed, quite generally, the momentum 
current tensor is not unique and the equilibirium condition~\eqref{eq:hydro} remains unchanged
under $(\underline{\Pi})^{ij}\to(\underline{\Pi})^{ij}+(\partial^i\partial^j-\delta^{ij}\nabla^2)\Phi$ with an arbitrary scalar function
$\Phi(x)$ which vanishes at infinity~\cite{nish07}. This freedom is removed by the concrete choice~\eqref{eq:hmf} below 
for $\mathcal{D}(x)$ in the central region of the self-bound droplet. \\

\end{itemize}

Explicit results for both the dipolar contact density $\mathcal{D}(x)$ and the local 
anomalous field $h(x)$ are available at the mean-field level. Specifically, we use
Eq.~\eqref{eq:hx} which connects the interaction contribution to $h(x)$ to the derivative 
of the dipolar contact density~\eqref{eq:Dx} with respect to the parameter $\theta$. 
Within mean-field, this yields 
 \begin{equation}
\mathcal{D}^{(\rm mf)}(x)=-\frac{1}{2}\langle\hat{M}_z(x)\rangle\,B_z^{\rm (mf)}(x)=-\frac{1}{2}g_{\rm dd}\,[\rho^{(1)}(x)]^2 f(\kappa) \; \to \; 
h^{(\rm mf)}(x)=-\frac{3}{4}g_{\rm dd}\,[\rho^{(1)}(x)]^2 b(\kappa)
\label{eq:hmf}
\end{equation}
where we have used the result~\eqref{eq:mf} for the magnetic field and the fact that the aspect ratio 
changes according to $d\kappa= -(3/2)\kappa\, d\theta$ under $x\to\theta[x]$. The positive function 
\begin{equation}
b(\kappa)=-\kappa\frac{df(\kappa)}{d\kappa} \;\to \;\left\{ \begin{array}{cc} 6\kappa^2\ln{(2/\kappa)} &  \mbox{for}  \;\; \kappa\ll 1 \\  
3\pi/(2\kappa) &  \mbox{for} \;\; \kappa\gg 1  
\end{array} \right.
\label{eq:bkappa}
\end{equation}
is determined by Eq.~\eqref{eq:demag} for the demagnetization tensor element $n^{(z)}$. 
It decays to zero for both strongly prolate or oblate configurations $\kappa\ll 1$ or $\kappa\gg 1$ and exhibits 
a broad maximum $b(\kappa=1.563\ldots)=0.852\ldots$ slightly above the value $b(\kappa=1)=4/5$ in the spherical limit.  
Now, the combination of Eq.~\eqref{eq:hmf} with the leading order result~\eqref{eq:psr} for the pressure due to 
short-range interactions shows that the two separate equilibrium conditions $p_{\rm sr}(x)+\mathcal{D}(x)=0$
and $h(x)=0$ for a self-bound droplet cannot be obeyed within mean-field where both $p_{\rm sr}(x)+\mathcal{D}(x)$
and $h(x)$ are negative for $\varepsilon_{\rm dd}\,f(\kappa)>1$ or arbitrary values of $\kappa$, respectively. A resolution of the first
problem, already proposed in the original publication on self-bound droplets~\cite{ferr16}, is provided by adding a
contribution to the short-range pressure which scales with a higher power in density than the quadratic behavior stated
as the low-density limit of the exact expression in Eq.~\eqref{eq:psr}. On a purely empirical level, a self-bound 
liquid can then be stabilized against collapse by extending the mean-field effective potential 
by a local term proportional to $\vert\psi(x)\vert^{2k}$ with $k>2$ in the form 
 \begin{equation}
 V_{\rm eff}[\psi]=-\mu\vert\psi(x)\vert^2+\frac{g}{2}\vert\psi(x)\vert^4+\frac{1}{2}\vert\psi(x)\vert^2 \int_{x'}V_{\rm dd}(x-x')\vert\psi(x')\vert^2+\frac{\lambda_k}{k}\vert\psi(x)\vert^{2k} \;\;\; (?)
 \label{eq:Veff}
 \end{equation} 
 where $\lambda_k>0$ is assumed to be positive. From a microscopic point of view,
 the true effective potential must, of course, be derived from the underlying Hamiltonian. 
 In fact, due to the anisotropy of the dipolar interaction, the simple form~\eqref{eq:Veff} where 
 the stabilizing contribution is rotation invariant, is not expected to be valid.  In principle, a microscopic derivation of
$ V_{\rm eff}[\psi]$ is possible by expressing the exact partition function in terms of a coherent state 
 functional integral~\cite{nege98}. The first three terms in Eq.~\eqref{eq:Veff} then arise 
 from minimizing the bare associated action. At this mean-field (or tree-) level, only
 two-body interactions enter. It provides a proper description of the zero density limit
 near the vacuum state $\psi(x)\equiv 0$, 
 where two-body scattering amplitudes are sufficient to deal with the many-body problem
 at finite density.  For self-bound droplets, which arise from the vacuum via a first-order 
 transition at a critical value $\mu_c<0$ of the chemical potential where the density jumps 
 from zero to a finite value $\bar{n}\ne 0$, this is no longer valid.  A formal procedure to extract 
 the form of $V_{\rm eff}[\psi]$ near the vacuum-to-liquid transition requires to 
 determine the generating functional $\Gamma[\psi_c]$ for the vertex functions from the exact 
 partition function $Z[J]$ in the presence of an external field $J$~\cite{itzy91,zee10QFT}. 
In practice, this has been achieved in Ref.~\cite{zwer19} for rotation invariant short-range 
interactions which become attractive beyond a zero crossing of the scattering length, where $g<0$.     
The exact effective potential obtained from $\Gamma[\psi_c]$ then indeed contains a cubic term $k=3$
as in~\eqref{eq:Veff} whose strength $\lambda_3=\hbar^2 D/(2m)$ is proportional   
the three-body scattering hypervolume $D$ introduced 
by Tan~\cite{tan08bose}. It has dimension (length)$^4$ and may be determined
for a given two-body interaction $V_{\rm sr}(r_{12})$ from a solution of the three-body scattering 
problem. Specifically, it is defined by the asymptotic behavior
\begin{equation}
\psi_{E=0}(x_1,x_2, x_3)\vert_{a=0}=1-\frac{\sqrt{3}\, D}{2\pi^3 (r_{12}^2+r_{13}^2+r_{23}^2)^2} \; +\ldots
\hspace*{0.3cm} {\rm if} \hspace*{0.3cm} V(x_{12})=V(r_{12})
 \label{eq:hypervolume}
 \end{equation} 
 of the three-body wave function at zero energy and vanishing scattering length. 
 The parameter $\lambda_3\sim D$ characterizes the strength of effective 
three-body interactions which arise beyond mean-field even for a microscopic Hamiltonian which only
contains two-body interactions. Explicit results for $D$ are available for  
simple model interactions like hard spheres~\cite{tan08bose}, an attractive square well~\cite{mest19} or a Lennard-Jones potential~\cite{mest20}.
In the two latter cases, the existence of two-body bound states gives rise to a finite imaginary part of the 
hypervolume. It determines the three-body loss coeffcient $L_3=-(\hbar/m)\,{\rm Im}\, D=(\hbar/m)\, L_{\rm rec}^4$ 
or the equivalent recombination length $L_{\rm rec}$~\cite{zhu17}. Now, for a possible stabilization of a liquid state
due to three-body interactions in the regime where $g_{\rm eff}(\kappa)=g-g_{\rm dd}f(\kappa)<0$ is negative,
it is the real part of the hypervolume $D$ which is relevant. In the realistic situation of finite three-body losses,
the parameter which determines the liquid density $\bar{n}=3|g_{\rm eff}|/(4\bar{\lambda}_3)$
from the condition $p_{\rm sr}(x)+\mathcal{D}(x)=0$ in the presence of
the additional pressure $\Delta p_{\rm 3-body}(x)=(2/3)\bar{\lambda}_3\,[\rho^{(1)}(x)]^{3}$ due to three-body interactions
is therefore $\bar{\lambda}_3=\hbar^2 \, {\rm Re}\,D/(2m)$.  
Evidently, ${\rm Re}\,D$ must be positive for a stable liquid, a condition which can be verified only by a concrete solution 
of the three-body problem. In particular, ${\rm Re}\,D$ assumes both positive and negative values, depending on
the position of three-body bound states which is determined by the poles of the hypervolume~\cite{zhu17,mest19,mest20}.  
In Ref.~\cite{ferr16} and the literature beyond, a stabilization of self-bound droplets by repulsive three-body interactions in 
the regime $g_{\rm eff}<0$ has been excluded by the argument that the experimental value of the loss 
coefficient $L_3$ gives rise to an estimate for the magnitude of the complex parameter $\lambda_3$ which is far too small to be consistent 
with the observed central droplet densities $\bar{n}\simeq 10^{-14}\, {\rm cm}^{-3}$. Since the real part $\bar{\lambda}_3\sim {\rm Re}\,D$
may be much larger than the estimate of $\lambda_3$ based on $|{\rm Im}\,D|$, this argument is not compelling, however. 
For the moment, we will postpone a further discussion of the origin of the beyond mean-field 
contribution $\Delta p\sim [\rho^{(1)}(x)]^{k}$ to the short-range pressure in Eq.~\eqref{eq:psr}
and proceed by assuming that such a term is present on purely empirical grounds. \\     
   
The shape-dependence of the effective coupling constant $g_{\rm eff}(\kappa)$ that determines the 
position of the mean-field instability gives rise to a corresponding shape-dependence 
of the liquid density, e.g. via $\bar{n}(\kappa)=3|g_{\rm eff}(\kappa)|/(4\bar{\lambda}_3)$ in the case 
of repulsive three-body interactions. For a homogeneous fluid, this is a rather strange conclusion and indeed
it is immediately changed if one takes into account that equilibrium in a uniaxial dipolar fluid requires in 
addition to $p_{\rm sr}(x)+\mathcal{D}(x)=0$ also that the anomalous field $h(x)=0$ must vanish.
Using the mean-field result~\eqref{eq:hmf}, this can be obeyed only if $b(\kappa)=0$ which implies that $\kappa$ 
is fixed to be either zero or infinity. The second possibility of an extremely oblate configuration is excluded since 
the dipolar interactions are  then repulsive and the zero pressure condition cannot be fulfilled. By contrast, the 
limit $\kappa\to 0$ of a strongly prolate droplet turns out to desribe the actual physical situation, consistent 
with the tendency of oriented dipoles to arrange in a head-to-tail configuration. More
precisely, the condition $h(x)=0$ gives rise to a finite value for the aspect ratio of a self-bound droplet rather 
than the singular result $\kappa^{({\rm mf})}\equiv 0$ within mean-field if one includes the contribution to $h(x)$ in 
Eq.~\eqref{eq:hx} which arises from the kinetic energy $\hat{H}_0$. This is a local term which - in contrast 
to the ill-defined kinetic contribution to ${\rm Tr}\,\underline{\Pi}(x)$ - is finite even if the range of $V_{\rm sr}(r_{12})$ is
taken to zero. Indeed, if the interactions become rotation invariant at very short distances, the contribution of large 
momenta in $h^{(0)}\sim \int_k (k_x^2+k_y^2 - 2k_z^2)\, n(k)$ cancels since the asymptotic momentum 
distribution $n(k)\to \mathcal{C}_2/|k|^4$ is then independent of the direction $\hat{k}$. A concrete result for
$h^{(0)}$ can be obtained by assuming an anisotropic Gaussian profile with central density $\rho^{(1)}(x=0)=\bar{n}$
\footnote{This assumption has been made in this context by Dalibard~\cite{dali24}. More precisely, an exact result due to 
Triay~\cite{tria19} shows that - within an extended Gross-Pitaevskii description and for a negative chemical potential - 
the density profile of a non-confined dipolar gas is smooth and falls off exponentially. }.            
In the relevant limit $\kappa\ll 1$, only the radial contribution in Eq.~\eqref{eq:hx} remains and the condition 
$h(x)=0$ in the center of the droplet is reduced to
\begin{equation}
\bar{n}\frac{\hbar^2}{2m}\pi\left(\frac{\bar{n}}{N\kappa}\right)^{2/3}=\frac{3}{4}g_{\rm dd}\bar{n}^2\cdot 6\kappa^2\ln{(2/\kappa)}
\hspace*{0.2cm} \to  \hspace*{0.2cm} \kappa^{8/3}\ln{(2/\kappa)}=\left[36\,\bar{n}^{1/3}a_{\rm dd}\cdot N^{2/3}\right]^{-1}=\epsilon\, .
 \label{eq:droplet}
 \end{equation} 
With typical particle numbers $N\simeq 10^4$ and central densities $\bar{n}^{1/3}a_{\rm dd}=\mathcal{O}(1)$, 
the smallness of the parameter $\epsilon\simeq 6\cdot 10^{-5}$ allows an asymptotically exact solution of the transcendental 
equation for $\kappa(\epsilon)$ in the form
 \begin{equation}
\kappa(\epsilon)=\left[\frac{\epsilon}{\ln{(2/\epsilon^{3/8})}}\right]^{3/8}
\hspace*{0.2cm} \to  \hspace*{0.2cm} \kappa(N)=\frac{0.26}{\left[\bar{n}^{1/3}a_{\rm dd}\cdot \ln{[2/\epsilon^{3/8}(N)]}\right]^{3/8}}\cdot N\,^{-1/4}\, .
 \label{eq:kappaN}
 \end{equation} 
 Apart from the logarithmic correction in the denominator which changes the result only by a numerical factor of order one,
 the aspect ratio thus approaches zero according to $ \kappa(N)\sim N^{-1/4}$ for large particle numbers. In practice, with 
 $N<10^5$, typical aspect ratios are in the range $1/\kappa\simeq 5 - 30$~\cite{bail16, bail17}. 
 An important point to note is that the result~\eqref{eq:kappaN} holds irrespective of the specific mechanism which 
 stabilizes the central density $\bar{n}$ at a finite value provided only that $\bar{n}^{1/3}a_{\rm dd}$ does not itself depend on $N$. 
 The dependence on the deviation $\varepsilon_{\rm dd}-1$ from the critical strength $\varepsilon_{\rm dd}^c=1$
 of the dipolar interactions, which is now shape-independent due to $\kappa\approx 0$, is sensitive to that, 
 however. In particular, for a beyond mean-field contribution $\sim\vert\psi(x)\vert^{2k}$, the liquid density approaches zero like
 $\bar{n}(\varepsilon_{\rm dd})\sim (\varepsilon_{\rm dd}-1)^{1/(k-2)}$. On physical grounds, the fact that the equilibrium
 configuration of self-bound dipolar fluids corresponds to an increasingly prolate object for large particle numbers is a 
 consequence of the anisotropy of the interactions. 
 In contrast to the situation of attractive short-range interactions stabilized by a repulsive three-body force 
 discussed in Refs.~\cite{zwer19,son20}, there is no homogeneous liquid state at all. This quite unusual 
 behavior is elucidated by a few further comments: 
 \begin{itemize}
\item The existence of a finite aspect ratio rather than the mean-field value zero relies on the kinetic
energy contribution to the anomalous field. This must be carefully distinguished from a 
standard quantum pressure term, which is able to stabilize a self-bound droplet by itself only in 1D.
Examples are the bright soliton in the Lieb-Liniger model with positive two-body scattering length $a_1>0$ or the droplets formed 
by attractive three-body interactions studied by Sekino and Nishida~\cite{seki18}. In both cases, the 
droplet extension approaches zero for large $N$ according to $R_N\to a_1/N$ or $R_N\sim\exp{[-(4/\sqrt{3}\pi) N^2]}$ 
instead of increasing like $L_z(N)\sim \sqrt{N}$ in the case of 3D dipolar gases.\\ 

\item There is a bound $N_0$ on the particle number below which the droplets evaporate. 
This has been investigated by Baillie et al.~\cite{bail16}, where $N_0(\varepsilon_{\rm dd})$ 
has been determined from an extended Gross-Pitaevskii description by the condition of a 
finite binding energy despite the competing quantum pressure term. The analogous 
problem for a Bose droplet with short-range interactions has been studied
in Ref.~\cite{zwer19} and - with concrete predictions for the lifetime of metastable droplets
in a finite window $N_1<N<N_0$  - by Son et al.~\cite{son22}. As discussed in Ref.~\cite{zwer19},
the critical number $N_0\sim \sqrt{D}/a^2$ in this context determines the scattering 
lengths at which $N$-body bound states detach from the continuum. \\ 

\item Non-spherical self-bound objects have been studied a long time ago
for stars with strong, frozen-in magnetic fields $B$ by Chandrasekhar and 
Fermi~\cite{chan53}. The balance between the decrease of 
the attractive gravitational energy with a finite excentricity and the gain in
magnetic energy due an expansion in the direction perpendicular to the field  
results in an oblate deformation with aspect ratio $\kappa -1\simeq (B/B_*)^2$,
 where $B_*$ is the field beyond which the star is no longer bound since magnetic pressure overwhelms gravitation. \\

 \end{itemize} 
 
 Following the suggestion in the original publication on self-bound droplets~\cite{ferr16}, 
 the mechanism for stabilizing a non-confined dipolar gas beyond the critical interaction 
 strength $\varepsilon_{\rm dd}=1$ where the homogeneous fluid is unstable is commonly 
 believed~\cite{chom23} to be a generalized form of the Lee, Huang and Yang (LHY) correction,
originally calculated for a dilute, homogeneous hard-sphere Bose gas~\cite{lee57}. 
 Including the effect of anisotropy in the presence of an additional dipolar interaction~\cite{lima11}, it gives rise 
 to a beyond mean-field contribution to the effective potential of the form introduced in Eq.~\eqref{eq:Veff} 
 with $k=5/2$ and a positive (as long as $\varepsilon_{\rm dd}\leq 1$) coefficient
 \begin{equation}
\lambda_{5/2}(g,\varepsilon_{\rm dd})=\frac{4g}{3\pi^2} (4\pi a)^{3/2}\!\cdot Q_5(\varepsilon_{\rm dd}) 
\hspace*{0.2cm} {\rm where} \hspace*{0.2cm} Q_5(\varepsilon_{\rm dd}) = \int_0^1\! dx\left[1+\varepsilon_{\rm dd}\cdot 2P_2(x)\right]^{5/2}\, .
 \label{eq:LHY}
 \end{equation} 
For the standard case of repulsive short-range interactions, the LHY term provides the leading correction beyond mean-field 
in the equation of state, giving rise to an additional contribution of order $gn^2\sqrt{na^3}$ to the pressure. It is important here to 
carefully separate the dependence on density from that on the chemical potential as the thermodynamically 
conjugate variable. In particular, for a homogeneous fluid, there is an exact relation $-V_{\rm eff}[n(\mu)]=p(\mu)$
which allows to infer the effective potential evaluated at the equilibrium value $n(\mu)=\partial p(\mu)/\partial \mu$
of the density from the pressure at given $\mu$~\cite{zwer19}. The increase of pressure with density which results from 
the positive LHY contribution in $V_{\rm eff}[\psi]$ thus arises from a corresponding negative correction to $p(\mu)$.
In explicit form, this emerges within a field-theoretic formulation where the beyond mean-field contribution determined 
by LHY is just the one-loop correction~\cite{zee10QFT,itzy91}  
  \begin{equation}
p^{(1)}(\mu)= -\frac{1}{2}\,{\rm Tr}\,\ln\,{\rm det}\,[\delta^{(2)}S]\; \xrightarrow[{\rm hom.}]\; -\frac{1}{2}\int_q \int_{\omega} \ln{[E_q^2+(\hbar\omega)^2]} 
\; \xrightarrow[{\rm reg.}]\; -\frac{8}{15\pi^2}\,\mu\left(\frac{m\mu}{\hbar^2}\right)^{3/2} 
 \label{eq:p1}
 \end{equation} 
to the tree-level result $p^{(0)}(\mu)=\mu^2/(2g)$ for the pressure of a dilute Bose gas
\footnote{Note that $p^{(1)}(\mu)$ does not explicitely depend on the scattering length. The contact density $\mathcal{C}_2(\mu)\to
(m\mu/\hbar^2)^2$ is therefore still given by its mean-field value and the LHY correction of relative order $(na^3)^{1/2}$, which
contributes to a corresponding one in the pressure~\eqref{eq:psr}, only shows up as a function of density.}. 
In physical terms, it describes the contribution to pressure due to zero-point fluctuations of the Bogoliubov excitations. Similar to the analogous 
Casimir effect, it is negative and it comes entirely from the low-energy part of the spectrum~\cite{zee10QFT,schw14QFT}. 
In particular, after regularization of the divergent integration over momenta $q$, the remaining expression
involves an integral $\int_q E_q^2\sim \int_0 q^2 dq\, c^2(\hat{q})q^2$. Here - in an extension to
anisotropic interactions - we include a possible dependence of the sound velocity $c(\hat{q})$ on direction. 
Now, it is a crucial point that the correction~\eqref{eq:p1} to the pressure is fully determined by the excitations near $q=0$. 
As a consequence, an anisotropic sound velocity $c^2(\hat{q})=c^2[1+\tilde{\varepsilon} f(\hat{q})]$ with 
strength $\tilde{\varepsilon}$ and an arbitrary function $f(\Omega)$ of direction changes
the LHY contribution~\eqref{eq:p1} just by a numerical factor $Q(\tilde{\varepsilon})=\int (d\Omega)/(4\pi)\,[1+\tilde{\varepsilon}f(\Omega)]^{5/2}$.
For the special case of dipolar interactions, this immediately explains the origin of the function $Q(\tilde{\varepsilon})\to Q_5(\varepsilon_{\rm dd})$ 
derived by Lima and Pelster~\cite{lima11}. It is obvious, however, that the LHY correction is well defined 
only as long as $c^2(\hat{q})>0$ remains positive. This requirement is violated in dipolar gases if $\varepsilon_{\rm dd}>1$,
where the homogeneous fluid is unstable.  
Formally, this shows up as a finite imaginary part of the function $Q_5(\varepsilon_{\rm dd})$ which is,
however, neglected in the extended Gross-Pitaevskii description of the inhomogeneous configurations
that are considered in practice. \\

For a proper understanding of the mechanism which underlies the stabilization of self-bound 
droplets and which allows to fulfill the zero pressure condition $p_{\rm sr}(x)+\mathcal{D}(x)=0$, it is of course
necessary to provide a physical argument that justifies the addition of a contribution $\Delta p_{\rm LHY}(x)=(3/5)\lambda_{5/2}\,[\rho^{(1)}(x)]^{5/2}$
of the LHY-form to the short-range pressure in a regime where the chemical potential is negative and the microscopic derivation
along the lines in Eq.~\eqref{eq:p1} evidently fails. On a rather qualitative level, the inclusion of an LHY term
even in the regime $\varepsilon_{\rm dd}>1$ may be justified by noting that the 
characteristic radial extension $R_{\perp}\sim N^{1/4}$ of self-bound droplets is only of the order of the healing 
length $\xi$~\cite{dali24}. As a result, there is no sound propagating in the $(x,y)$-plane in such a configuration 
and the fact that $c^2(\hat{q})<0$ for the associated wave vectors is irrelevant. The argument is not conclusive, however, because it does
not address the problem that an exponent $5/2$ in the relation between pressure and density requires a corresponding
power law in $p(\mu)$, which is excluded for negative values of $\mu$. To obtain some further insight, it is useful 
to investigate the microscopic origin of the LHY correction from a different point of view. The associated characteristic exponent $k=5/2$ 
indicates that it can be attributed neither to two-body nor to three-body interactions which - as discussed above - 
lead to $k=3$. Now, as noted already by Lee, Huang and Yang~\cite{lee57} and expanded in detail later by
L\"uscher~\cite{lues86} and by Tan~\cite{tan08bose}, the LHY term can be understood to arise from a finite size correction
in the two-body problem. More precisely, the repulsive contribution $\sim gn\, (na^3)^{1/2}$
in the interaction energy per particle beyond mean-field emerges from a correction of order $a/L$
beyond the leading term $E^{(0)}_{N=2}(L)=g/L^3$ in the two-particle problem in a box with periodic boundary conditions.
At finite density, the relevant size $L\simeq \xi$ for this correction is set by the healing length beyond which 
the pair distribution is no longer determined by two-body physics. In the presence of the long-range and anisotropic
dipolar interactions, both $E^{(0)}_{N=2}(L)$ and its finite size correction will depend on the
aspect ratio $\kappa$ and the dimensionless strength parameter $\varepsilon_{\rm dd}$. A 
microscopic derivation of the LHY correction for dipolar gases in a configuration which mimics
the elongated droplets thus requires the solution of the two-body problem in an anisotropic box of 
size $L\times L\times L/\kappa$ with periodic boundary conditions. Specifically, the parameter
$\lambda_{5/2}(\kappa,\varepsilon_{\rm dd})$ is determined 
by the coefficient of the $a/L$ - contribution to the energy in a finite size expansion analogous 
to that derived by L\"uscher for rotation invariant short-range interactions~\cite{lues86}. While  
straightforward in principle, proving that $\lambda_{5/2}$ is finite and positive along these lines is 
nontrivial. A fully microscopic derivation of either
the LHY correction or a possible repulsive three-body force that might contribute to the 
stabilization of self-bound droplets thus remains an open problem. It should be emphasized,
however, that this is essentially a conceptual challenge, not a practical one. Indeed, as 
demonstrated recently by Bomb\'{i}n et al.~\cite{bomb24}, numerical approaches to dipolar
gases in the regime $\varepsilon_{\rm dd}>1$ based on quantum Monte Carlo provide results which agree quite well 
with experiment, e.g. for the critical number $N_0$ where droplets unbind or for the supersolid transition.\\

\section{Effective theories for the transition to a supersolid}
The prediction that dipolar gases in a pancake geometry with $\lambda \gg 1$ will 
exhibit a roton-maxon character in their excitation spectrum has been made even 
before the experimental realization of dipolar condensates~\cite{sant03,dell03}. 
In contrast to the situation in $^4$He, where the roton gap $\Delta_r$ is 
essentially independent of pressure up to the superfluid-to-solid transition at $p_c\simeq 25\, {\rm atm}$, 
the depth of the roton minimum is now tunable. In fact, within mean-field, the roton gap 
in dipolar gases is found to vanish beyond a critical value of the interaction strength where 
the spectrum of Bogoliubov excitations reaches zero~\cite{sant03,dell03}.
A simple model capturing the essential physics is based on assuming a uniform quasi-2D situation
with a Gaussian density profile in the transverse direction with characteristic length $\ell_z$.  
The resulting effective dipole-dipole interaction~\cite{dell03,fisc06dipolar} 
\begin{equation}
V_{\rm dd}(q) = \frac{\hbar^2}{m}\tilde{g}_{\rm dd}\left[2-3 \sqrt{\frac{\pi}{2}} (q\ell_z) \exp{(q^2\ell_z^2/2)}\,{\rm erfc}(q\ell_z/\sqrt{2})\right].  
\label{eq:Fischer}
\end{equation}
in momentum space $q=(q_x,q_y)$ approaches a positive constant $V_{\rm dd}(q) \to 2\hbar^2\tilde{g}_{\rm dd}/m$ in the limit  $|q|\ell_z \ll 1$,
where $\tilde{g}_{\rm dd}=\sqrt{8\pi} a_{\rm dd}/\ell_z$ is a dimensionless coupling constant.
The fact that $V_{\rm dd}(q=0)$ is positive guarantees that - in contrast to the situation in 3D - the homogeneous fluid 
is now stable for an arbitrary strength of the  dipolar interaction. This does not exclude an instability at finite wave vector, however.  
Indeed, the effective interaction $V_{\rm dd}(q)$ turns negative beyond $|q|\ell_z$ of order one and asymptotically 
approaches the constant value $-\hbar^2\tilde{g}_{\rm dd}/m$. In physical terms, this describes attractive head-to-tail 
collisions between aligned dipoles at distances less than $\ell_z$ with an effective scattering length $-a_{\rm dd}$.
For strong dipolar interaction strengths, the negative contributions to the total interaction 
$V(q)=\hbar^2\tilde{g}_2/m +V_{\rm dd}(q)$ give rise to a roton minimum in the spectrum $E_q^2=2nV(q)\varepsilon_q+\varepsilon_q^2$
of Bogoliubov excitations which eventually touches zero at a characteristic wave vector  $q_0$. A crucial point in this context is
that the scale for $q_0$ is set by the inverse of the confinement length $\ell_z$, which is unrelated to and much larger than
the average interparticle spacing in the transverse direction.\\
  
The predicted appearance and subsequent softening of a roton has been observed in dipolar condensates 
of Dysprosium in a cigar-shaped trap by Petter et al.~\cite{pett19} using Bragg spectroscopy. By tuning $\varepsilon_{\rm dd}=a_{\rm dd}/a$
via a Feshbach resonance that allows to change the short-range scattering length at fixed $a_{\rm dd}$, the measured 
dynamic structure factor shows a sharp increase near a critical value $\varepsilon_{\rm dd}^c$ of
order one, consistent with theoretical results based on a solution of the Bogoliubov equations~\cite{blak12}.
This signals a macroscopic occupation of the roton mode associated with a spontaneous density modulation
along the weakly confined direction in the trap, as seen directly from in-situ density profiles~\cite{hert20}.
The formation of a static density wave at a coupling strength where the roton dip is still not very pronounced 
and the interpretation of this state in terms of a supersolid has been confirmed in a number of 
experiments~\cite{chom23} which will not be discussed here in detail. Instead, we will present 
an elementary approach which allows to understand the nature of the associated phase transition and the 
underlying physics independent of specific details. In particular, based on a simple model due to Nozi\`eres~\cite{nozi04}, 
we will show that the mean-field roton instability is preempted into a first-order transition by an amount which depends 
on the strength of the short-range repulsion. As a first step, it is necessary to properly define the notion of a 
 supersolid which, in a rather broad sense, may be characterized by:\\
 
{\bf In a supersolid, superfluidity} (defined by a finite superfluid fraction)
{\bf appears together with a non-vanishing modulation of the density due to spontaneously broken translation invariance}. \\

Now, according to this definition, any spatially modulated superfluid like the vortex lattice is 
also a supersolid. To exclude such well known cases, the notion of a genuine supersolid should 
therefore be restricted to phases where spontaneously broken translation invariance
and superfluidity are present simultaneously as {\it two independent} order parameters.
The modulation in density is then still present even after superfluidity is lost.
An important result due to Leggett~\cite{legg70} states that in any superfluid 
with a non-uniform density, there is an upper bound on the superfluid fraction $f_s$ 
strictly smaller than one. Specifically, the Leggett bound is of the form 
 \begin{equation}
\hspace*{-0.1cm} f_s(T=0)\leq \left[ \frac{\bar{n}}{d} \int_0^d \frac{dx}{n_1(x)} \right]^{-1}\; ,  \hspace*{0.3cm} {\rm in\;mean-field} \hspace*{0.3cm} f_s^{({\rm mf})}(T=0)=m/m_B
  \label{eq:Leggett}  
 \end{equation}
where the inverse of the density $n_1(x)$ is integrated over a unit cell of the lattice whose
length is denoted by $d$. Here, $n_1(x)$ is the average of the microscopic density $\langle\hat{\rho}^{(1)}(x)\rangle$ 
over the transverse directions of the unit cell and $\bar{n}$ its average value (without loss of generality, 
the $x$-direction has been singled out). Apparently, the 
bound is always finite unless the density vanishes identically in some region. Within mean-field, 
the ground state wave function factorizes into a product of single-particle ones. The Leggett bound 
then turns into an equality and relates the superfluid fraction to the ratio $m/m_B$ of the bare and the band mass $m_B$
in the given periodic potential. For a conventional solid, where the particle density is 
concentrated near a discrete set $\{\mathbf{R}\}$ of lattice sites with an exponentially suppressed value at 
interstitial positions, the upper bound on $f_s$ is much smaller than one. In the limit of a fluid with uniform 
density, in turn, Eq.~(\ref{eq:Leggett}) reduces to the trivial identity $f_s\leq 1$. Obviously, supersolids with an 
appreciable superfluid fraction can only be found in situations where the density modulation is weak. A few 
important points should be noted in this context:
 First, the bound~(\ref{eq:Leggett}) does not provide a sufficient criterion for superfluidity in a state with broken 
 translation invariance: a finite value of the bound is still compatible with no superfluidity at all. A case in point 
 is the solid phase of $^4$He, where the microscopic density profile is known from path integral Monte Carlo 
 and the Leggett bound gives $f_s\leq 0.16$~\cite{cepe04} while the true value vanishes
 \footnote{For a discussion of why a supersolid phase of bulk $^4$He can be ruled out, see Refs.~\cite{cepe04,boni06} and the review~\cite{boni12rmp}.}.
 A second point is that 
 the bound~(\ref{eq:Leggett}) makes no assumption about the physical origin of the density modulation. It may 
 arise from a spontaneous breaking of translation invariance but it also holds if the density modulation is 
 externally imposed. In fact, the latter method has been used in a recent experimental test of the Leggett bound~\cite{chau23}. 
 Finally, a quite subtle point is that the bound holds independent of whether the number of particles within a unit cell happens to be an 
 integer or not. Now, as emphasized in the classic paper by Andreev and Lifshitz~\cite{andr69}, the generic 
 realization of supersolids requires an incommensurate situation with a finite concentration of defects in the
 ground state. A detailed argument which shows that delocalized vacancies or interstitial atoms are indeed a 
 necessary condition for a supersolid has been given by Prokof'ev and Svistunov~\cite{prok05supersolid}. 
 This suggests that supersolids are in general just superfluid mass-density waves, with the homogeneous 
 part of the density playing the role of delocalized 'defects'. The ground state of generic solids like $^4$He, in turn, has an 
 integer number of particles per unit cell and vanishing defect concentration~\cite{boni06}.The finite energy necessary for the 
 creation of either vacancies or interstitials identifies such a state as a Mott-insulator, which cannot have a finite superfluid fraction. \\

\subsection{Mean-field theory of freezing in classical and quantum systems}

It is a classic argument due to Landau~\cite{land37} that the transition from a uniform fluid to a 
state with a periodic modulation of the density is of first order. The argument relies on an 
expansion of the free energy in terms of the Fourier components $n_{\mathbf{G}}\ne 0$ of the density
\begin{equation}
\rho^{(1)}(x)=\langle \rho^{(1)}(x)\rangle +\sum_{\mathbf{G}\ne 0}n_{\mathbf{G}}e^{i\mathbf{G}\cdot\mathbf{x}}=n_{\rm sol}+\sum_{\mathbf{G}\ne 0}n_{\mathbf{G}}e^{i\mathbf{G}\cdot\mathbf{x}}
 \label{eq:Fourier}
 \end{equation} 
in the symmetry broken phase with average density $n_{\rm sol}$, where $\{\mathbf{G}\}$ denotes the set of associated reciprocal lattice vectors~\cite{chai95}. 
In the weak crystallization limit, the density jump $\Delta n=n_{\rm sol}-n_{\rm liq}$ is small and the instability is dominated by 
a single wave vector $q_0$ where the static structure factor of the homogeneous fluid exhibits a pronounced maximum $S(q_0)$.
The magnitude $|\mathbf{G}|=q_0$ of the reciprocal lattice vectors is then fixed and the problem reduces to 
finding the lattice with the lowest free energy. Within Landau theory, this is of the generic form~\cite{chai95}
 \begin{equation}
\Delta f_L(T)=f_{\rm sol}-f_{\rm liq}=\frac{r(T)}{2} n_G^2 - w n_G^3+u n_G^4 \hspace*{0.2cm} {\rm with} \hspace*{0.2cm} r(T)=T-T^*  \hspace*{0.2cm} {\rm and} \hspace*{0.2cm}  u>0\, .
 \label{eq:Landau}
 \end{equation} 
 Here, $n_G^2 \sim \sum_{\mathbf{G}}' |n_{\mathbf{G}}|^2$ is a sum of the magnitudes of the non-vanishing 
 Fourier components with fixed $|\mathbf{G}|=q_0$, rescaled in such a way that makes it independent of
 their number in the specific lattice and dimensionless
 \footnote{Note that the sign of $w$, assumed to positive here, is irrelevant since it can be 
 changed trivially by $n_G\to -n_G$.}.
 Now, the simplified form~\eqref{eq:Landau} hides the dependence on the specific lattice structure
 which is contained in the detailed values of $w$ and $u$. For a finite $w\ne 0$, reciprocal lattices in 
 which triads of different ${\mathbf{G}}'$s add up to zero are favored. At fixed $|\mathbf{G}|=q_0$,
 therefore, the planar hexagonal, the fcc and icosahedral lattices are the only possibilities~\cite{chai95},
 giving rise to a triangular lattice in real space as the unique option in 2D. Quite generally, 
the transition temperature $T_c$ is determined by the condition 
 $\Delta f_L(T_c)=0$ at a finite value $n_G(T_c)=w/(2u)\ne 0$, where the fluid with $n_G\equiv 0$ is 
 degenerate with a state where the density develops a non-vanishing spatial modulation. Due to the 
 presence of the third-order term, the critical temperature $T_c=T^*+w^2/(2u)$ lies above 
 the temperature $T^*$ introduced in the phenomenological Ansatz for $r(T)$. Physically, $T^*$ is
 the lowest temperature down to which the fluid exists as a metastable configuration. Despite its 
 purely phenomenological nature, the Landau theory of crystallization captures many aspects of real
 first-order fluid-to-solid transitions, like the preference for (real space) bcc lattices near the melting line~\cite{alex78}
 or the fact that a fluid can be undercooled by a far larger amount than solids can be overheated. Indeed,
 the temperature below which the solid exists at least as a metastable configuration is given 
 by $T_1=T^*+9w^2/(16u)$ and the Ansatz in Eq.~\eqref{eq:Landau} thus leads to 
 $T_c-T^*=8\,(T_1-T_c)$.   A further consequence of Landau theory, not mentioned usually, is that the static structure 
 factor in the uniform fluid right at the transition has a quasi-universal value $S(q_0)\vert_{T_c}=T_c/r(T_c)=T_c/(T_c-T^*)$
 of order one. A similar result was in fact found to hold for strong crystallization transitions e.g. of a Lennard-Jones
 fluid by Hansen and Verlet~\cite{hans69}. It has led to the empirical Hansen-Verlet criterion $S_{\rm HV}(q_0)\vert_{T_c}=2.85$
 as a system-independent estimate for the position of the fluid-to-solid transition in classical systems with short-range 
 interactions from properties within the fluid phase itself.          \\
 
 The Landau theory for the fluid-to-solid transition in a classical system can be extended 
 into a quantum theory which describes the appearance of a mass-density wave in a superfluid. 
 Based on an approach indicated already in the ground-breaking work of Gross on inhomogeneous
 Bose fluids~\cite{gros58}, this has been developed by Pomeau et al.~\cite{pome94,joss07}.
 Focussing again on density fluctuations with wave vectors of a fixed length $|\mathbf{G}|=q_0$, the leading 
 order expansion $\sum_q |\delta n_q|^2/2\chi(q)$ in the change of energy through density fluctuations $\delta n_q\ne 0$
in any fluid gives rise to a contribution $\tilde{r}n_G^2/2$, where $\tilde{r}=1/\chi(q_0)$ is the inverse of the static 
 density response function. Now, within the Bogoliubov approximation,
 this response is determined by the ratio $\chi(q)=2\varepsilon_q/E_q^2$ of the free particle energy $\varepsilon_q$ 
 and the dispersion $E_q$ of the collective excitations of the superfluid~\cite{pita16}. In particular, for interactions that lead to
 a roton minimum with energy $\Delta_r=E_{|q|=q_0}$, the parameter $\tilde{r}$ near the corresponding wave vector
 is given by $\tilde{r}=\Delta_r^2/(2\varepsilon_r)$ with $\varepsilon_r=\varepsilon_{|q|=q_0}$. 
 It is important to note that -  within the Bogoliubov approximation - a roton minimum requires two-body
 interactions $V(q)$ which are negative in a certain range of wave vectors but not necessarily an attractive interaction
in real space as in dipolar gases. As an example, a purely repulsive box potential $V(r_{12})$ of strength $V_0$ and size $\sigma$ 
 gives rise to a negative $V(q)$ in a range $4.5<q\sigma< 7.7$. The energy of the resulting roton minimum and the associated
parameter $\tilde{r}$ can then be tuned by changing the dimensionless coupling strength $n\sigma^3\cdot V_0\, m\sigma^2/\hbar^2$.  
This leads to a roton instability at the point where $\tilde{r}$ vanishes and the uniform superfluid becomes unstable. 
The nontrivial question now is whether this instability can be cured by a nonlinear contribution to the energy 
analogous to the one which is introduced by hand in Landau's Ansatz~\eqref{eq:Landau}. As shown by
Pomeau and Rica~\cite{pome94} this is indeed the case and the resulting energy functional   \\       
  \begin{equation}
\Delta f_{\rm GP}=f[n(x)]-f[\bar{n}]=\frac{\tilde{r}}{2} n_G^2 - \tilde{w} n_G^3+\tilde{u} n_G^4 \hspace*{0.2cm} {\rm with} \hspace*{0.2cm} \tilde{r}=\Delta_r^2/(2\varepsilon_r)
 \label{eq:Pomeau}
 \end{equation} 
 is in fact of the same form. Moreover, in contrast to the case of classical fluid-to-solid transitions where the parameters $r(T), w, u$ 
 are introduced in a purely phenomenological manner, the coefficients in Eq.~\eqref{eq:Pomeau} can now be 
 derived from a microscopic energy functional. In particular, the nonlinear terms involving $\tilde{w}$ and $\tilde{u}$ follow from the 
 expansion of the denominator in the quantum pressure contribution $\hbar^2\,[{\rm grad}\, n(x)]^2/[8m\, n(x)]$ around the fluid 
 state with uniform density $\bar{n}$. For a given characteristic wave vector $q_0$ of the emerging density wave, they are on
 the order of the associated recoil energy $\tilde{w},\tilde{u}\simeq \varepsilon_r$, with detailed values again 
 depending on the specific lattice. As in the standard Landau theory, the presence of a finite third-order term $\tilde{w}\ne 0$ 
 implies that the uniform superfluid freezes with a jump in density determined by $n_G\vert_c=\tilde{w}/(2\tilde{u})$ at a critical value $\tilde{r}_c=\tilde{w}^2/2\tilde{u}$. 
 The mean-field roton instability at $\tilde{r}=0$, which characterizes the interaction strength up to which the 
 homogeneous superfluid exists at least as a metastable configuration, is thus preempted by first-order transition 
 at a finite critical value $\Delta_r/\varepsilon_r\vert_c=\mathcal{O}(1)$ of the roton gap. For the specific case of 
a triangular lattice in 2D, this value turns out to be $\Delta_r/\varepsilon_r\vert_c=0.23$~\cite{pome94}.    \\
 
 \subsection{Effects beyond mean-field and a two-mode model}
 
The extension above of Landau's classical theory to the freezing of a uniform superfluid into a 
state with broken translation invariance describes a mean-field supersolid in the sense 
that its condensate fraction stays at the non-interacting value $f_0^{({\rm mf})}=1$.
By contrast - consistent with the Leggett bound~(\ref{eq:Leggett}) - the superfluid fraction 
is reduced below one depending on the magnitude of the density modulation.
The model thus provides a qualitative description of the supersolid transition in dipolar 
gases where the emerging density wave is a phase-coherent superfluid. This is
rather different from the situation in $^4$He, where the solid phase beyond the 
critical pressure $p_c$ is a commensurate Mott-insulator~\cite{boni06}. 
The result that the roton gap has a finite value $\Delta_r^c$ at the supersolid transition 
implies a quantum generalization of the Hansen-Verlet criterion. Indeed, the expression $\chi(q)=2\varepsilon_q/E_q^2$ 
for the density response also holds for finite temperatures below the superfluid transition 
as long as the condensate fraction remains close to one~\cite{pita16}. As a result,
the static structure factor of the homogeneous superfluid at finite temperature is related to 
its ground state value $S(q)=\varepsilon_q/E_q$ by a thermal enhancement 
factor $\coth{(E_q/2T)}$. Right at the supersolid transition, it has thus a quasi-universal value 
 \begin{equation}
S_c(q_0)=\left(\frac{\varepsilon_r}{\Delta_r^c}\right)\,\coth{\left(\frac{\Delta_r^c}{2 T}\right)} \;\; \xrightarrow[T\gg\Delta_r^c]\;\; \frac{T}{\tilde{r}_c}\gg 1
 \label{eq:HV}
 \end{equation} 
 determined only by the ratios between the roton gap $\Delta_r^c$ and the recoil energy or temperature.
 As indicated, the value is large compared to one in the experimentally relevant limit $T\gg\Delta_r^c$. 
 In particular, this applies to the measurements of the static structure factor in Ref.~\cite{hert20}, where 
 a pronounced peak in $S(q_0)$ near the ordering wave vector is found at the supersolid transition.\\  

There are two major shortcomings of the approach sketched above:
First of all, the description fails in cases where the emerging lattice does not give
rise to a third-order invariant in the free energy. This applies e.g. to the situation
realized in many experiments where the density modulation appears along a single 
direction preferred by the geometry in cigar-shaped traps. 
The functional~\eqref{eq:Pomeau} then leads to a continuous transition at the point 
where the roton gap vanishes which is not consistent with observation
\footnote{This problem may be eliminated by incorporating a further wave vector $q_0'$ for 
the density wave, which breaks the symmetry $n_G\to -n_G$. The option comes, however, at the 
expense of introducing at least two additional free parameters.},
see e.g. Ref.~\cite{hert20}.  A second point is that, 
even at finite values of $\tilde{w}$, the shift of the roton instability away from its 
mean-field value zero to a finite $\tilde{r}_c=\tilde{w}^2/2\tilde{u}$ predicted by Eq.~\eqref{eq:Pomeau} 
gives rise to a critical roton gap of order $\varepsilon_r$, independent of the strength of interactions. 
In the following, we will outline an approach in which both of these problems are absent.
The crucial point to recognize is that beyond mean-field there is a 
fundamental difference between density fluctuations and the quasi-particles of the superfluid.
It is only in the limit of weak interactions where they coincide and one may infer the  
density response $\chi(q)\to 2\varepsilon_q/E_q^2$ from the quasi-particle dispersion $E_q$. 
A simple model for dealing with the interplay of density modes and superfluid 
quasi-particles was suggested by Nozi\`eres~\cite{nozi04}, whose aim was to shed light
on the complex question about the role of the roton minimum in superfluid $^4$He
for the transition to a solid. Based on a diagrammatic argument, Nozi\`eres showed 
that beyond the standard leading order term $\Sigma_q(\omega)\to n_0\, V(q)$ of the self-energy,
there is a contribution proportional to $V^2(q)\cdot\chi(q,\omega)$ which involves the 
dynamic density response function $\chi(q,\omega)$. Within a simple Ansatz $\chi(q,\omega)\simeq 2\varepsilon_q/(\Omega_q^2-\omega^2)$
for this response in the absence of the coupling to superfluid quasi-particles, a two-mode model emerges whose excitation energies $E_{\pm}^2(q)$
follow from a bi-quadratic equation ~\cite{nozi04}
 \begin{equation}
E_{\pm}^2(q)=E_q^2-f_0\,\frac{\Lambda_q}{\Omega_q^2- E_{\pm}^2(q)} \hspace*{0.1cm} \to  \hspace*{0.1cm}
\tilde{r}^{\rm eff}(q)=\frac{E_{-}^2(q)}{2\varepsilon_q}\simeq \frac{E_q^2}{2\varepsilon_q}-f_0\,\frac{\Lambda_q}{2\varepsilon_q\cdot E_{q}^2}=
\tilde{r}^{\rm mf}(q)-\frac{\, f_0\Lambda_q/(2\varepsilon_q)^2}{\tilde{r}^{\rm mf}(q)}
 \label{eq:modes}
 \end{equation} 
 The effective coupling contains the condensate fraction $f_0$ and the parameter $\Lambda_q$ 
 which is at least quadratic in the interactions. With increasing coupling, the two 
 modes $E_{\pm}^2(q)$ shift in an opposite direction and it is the lower one $E_{-}(q)$ whose softening 
 signals the onset of a density wave at given $q$. In particular, the associated effective stiffness $\tilde{r}^{\rm eff}(q)$ 
is renormalized down compared with the mean-field value by an amount which scales inversely
with $\tilde{r}^{\rm mf}(q)$ itself
\footnote{This is reminiscent of the Brazovskii equation in the beyond mean-field description of classical fluid-to-solid
transitions~\cite{braz75,braz87} but note that the sign here is opposite and the effect scales with the condensate fraction $f_0$.}.  
 The fact that a mean-field approximation overestimates the stiffness for density fluctuations 
 is consistent with the exact inequality $m_p^2(q)\leq m_{p+1}(q)\, m_{p-1}(q)$ for the 
 $p$-th moment of the dynamic structure factor~\cite{pita16}. In fact, in the special case $p=0$, this reduces to  
 \begin{equation}
S^2(q)\leq \varepsilon_q\cdot \chi(q)/2  \hspace*{0.2cm} \to  \hspace*{0.2cm}
1/\chi(q)\leq\frac{\varepsilon_{q}}{2S^2(q)}  \;\; \xrightarrow[{\rm mf}]\;\;\frac{\Delta_r^2}{2\varepsilon_r}
 \label{eq:Price}
 \end{equation} 
in general, respectively evaluated at the roton minimum within mean-field. It is only in the limit where the dynamic 
structure factor has a single sharp peak that Eq.~\eqref{eq:Price} becomes an equality.  A quantitative result 
for the renormalization of $\tilde{r}^{\rm mf}(q)$ requires to determine the parameter $\Lambda_q$. 
This may be estimated by assuming that the relevant wave vector is large 
enough that $\Lambda_q$ can be inferred from the exact short-distance expansion
 \begin{equation}
\frac{1}{\chi(q)}\;\to\; \frac{\varepsilon_q}{2}\left[ 1- \frac{\pi\mathcal{C}_2}{8nq}+\ldots\right] \hspace*{0.3cm} \to  \hspace*{0.3cm}
\frac{\Lambda_q}{\varepsilon_q^4}=\frac{\pi\mathcal{C}_2}{8nq} 
 \label{eq:OPE}
 \end{equation} 
for Bose gases with repulsive short-range interactions derived in Ref.~\cite{hofm17bragg}. 
The requirement that this is consistent with Eq.~\eqref{eq:modes} then fixes $\Lambda_q\sim\varepsilon_q^4\, \mathcal{C}_2/q$
to be proportional to the two-body contact density $\mathcal{C}_2$. Now, the characteristic wave vectors  
of the actually observed supersolids are considerably smaller than those where the expansion for high momenta in~\eqref{eq:OPE}
is expected to apply and - moreover - effects of the long-range dipolar interactions have been ignored. The result is
thus only of a rather qualitative nature. Nevertheless, it indicates how the problems of the mean-field approximation 
mentioned above can be resolved. Specifically, from the fact that the inverse static density response $\tilde{r}^{\rm eff}(q_0)$
of the homogeneous fluid at the freezing transition is expected to be finite but very small, the critical value         \\
 \begin{equation}
\left(\frac{\Delta_r}{\varepsilon_r}\right)_c^4\simeq f_0\, \frac{\pi\mathcal{C}_2}{8nq_0} 
 \label{eq:ratio}
 \end{equation} 
 for the ratio between the roton gap $\Delta_r$ and the recoil energy $\varepsilon_r$ at the transition into 
 a supersolid state can be estimated. Independent of the existence of a third-order invariant, it is
 always finite. Moreover, it depends explicitely on the interaction strength via the two-body contact density $\mathcal{C}_2$.
 In particular, the simple mean-field roton instability with a vanishing $\Delta_r^c\vert^{({\rm mf})}\equiv 0$ is obtained
 in the limit of vanishing short-range repulsion, where $\mathcal{C}_2\sim a^2\to 0$. It is obvious that the rather crude arguments 
 above need to be replaced by a proper microscopic theory for effects beyond mean-field in the transition to a supersolid.
 This is an open but clearly quite challenging problem.\\ 
  
\section{Conclusion and open problems}
The thermodynamic approach developed in this work provides a
description of dipolar quantum fluids which fully accounts for the anisotropy
and the long-range nature of the interactions. It allows to derive a number of
exact results, e.g. for the effective magnetic field in the center of a spherical cloud~\eqref{eq:K}
and the resulting instability beyond a critical value of the interaction strength,  
the violation of the Gibbs-Duhem relation in the density profile of trapped gases~\eqref{eq:GD} 
or the number-dependence~\eqref{eq:kappaN} of the aspect ratio in a self-bound droplet. 
An interesting perspective to obtain further exact results is provided by a possible extension
into an effective field theory, similar to the one that has been developed by Son 
and Wingate for the unitary Fermi gas~\cite{son06}. Concerning the transition into 
a supersolid phase, the discussion in this work has been of a rather qualitative nature
and a better understanding of the effects beyond mean-field is an open problem. \\

It is obvious, that a number of basic problems in the theory of dipolar gases have only been raised but have 
not been answered. This is true, in particular, for a microscopic derivation of the exact effective potential 
which replaces the simple ad hoc Ansatz assumed in Eq.~\eqref{eq:Veff}.  Specifically, both the detailed  form 
of an LHY-like contribution $\sim |\psi|^5$ with a proper positive strength $\lambda_{5/2}(g,\varepsilon_{\rm dd},\kappa)>0$
and a quantitative theory for a possible contribution of three-body forces are missing. Moreover, the transition from the supersolid phase to a
collection of separate, incoherent droplets~\cite{chom23} is still not well understood. \\  

Finally, the approach to uniaxial quantum fluids presented here applies to a situation where the breaking of
rotation invariance is imposed externally by the fixed orientation of the dipoles. It 
is a challenge to see whether it can be extended to cover a situation where the
symmetry breaking is spontaneous, e.g. in a quantum version of a nematic liquid crystal
\footnote{I am grateful to Alan Dorsey for pointing out such a possibility.}.   \\

{\bf Acknowledgement}: The present contribution is based on a Lecture on dipolar gases and supersolids given at 
the Institut Henri Poincar\'{e} in July 2024. I am very grateful to Yvan Castin for the invitation to this Lecture
and for extensive discussions. It is also a pleasure to thank Jean Dalibard for a number of comments 
and a copy of the Notes for his Lectures on the subject at the Coll\`ege de France~\cite{dali24}. Finally, I am 
indebted to Markus Holzmann for clarifying some basic points about the freezing transition of quantum fluids.\\

\bibliographystyle{crunsrt}

\nocite{*}

\bibliography{dipolar-bib}

\end{document}